\documentclass[12pt]{article}

\usepackage{latexsym,amsmath,amscd,amssymb,graphics}
\usepackage{enumerate,framed}

\usepackage{graphicx}

\usepackage[square,authoryear]{natbib}
\usepackage[colorlinks]{hyperref}
\usepackage{url}

\usepackage[all]{xy}

\newcommand{\PCF}{$\operatorname{PCF}$}
\makeatletter

\@addtoreset{figure}{section}
\def\thefigure{\thesection.\@arabic\c@figure}
\def\fps@figure{h, t}
\@addtoreset{table}{bsection}
\def\thetable{\thesection.\@arabic\c@table}
\def\fps@table{h, t}
\@addtoreset{equation}{section}

\makeatother

\begin{document}

\newtheorem{theorem}{Theorem}[section]
\newtheorem{definition}[theorem]{Definition}
\newtheorem{lemma}[theorem]{Lemma}
\newtheorem{remark}[theorem]{Remark}
\newtheorem{proposition}[theorem]{Proposition}
\newtheorem{corollary}[theorem]{Corollary}
\newtheorem{example}[theorem]{Example}

\def\below#1#2{\mathrel{\mathop{#1}\limits_{#2}}}



\title{Variational principles for spin systems and the Kirchhoff rod}
\author{Fran\c{c}ois Gay-Balmaz$^{2}$, Darryl D. Holm$^{1}$  and Tudor S. Ratiu$^{2}$}
\addtocounter{footnote}{1}
\footnotetext{Department of Mathematics, Imperial College London. London SW7 2AZ, UK. Partially supported by Royal Society of London Wolfson Award.
\texttt{d.holm@imperial.ac.uk}
\addtocounter{footnote}{1} }
\footnotetext{Section de
Math\'ematiques and Bernoulli Center, \'Ecole Polytechnique F\'ed\'erale de
Lausanne.
CH--1015 Lausanne. Switzerland. Partially supported by a Swiss NSF grant.
\texttt{Francois.Gay-Balmaz@epfl.ch, Tudor.Ratiu@epfl.ch}
\addtocounter{footnote}{1} }

\maketitle

\makeatother

\maketitle


\noindent \textbf{AMS Classification:} 

\noindent \textbf{Keywords:} Affine Euler-Poincar\'e equations, Lagrangian reduction, Clebsch-constrained variational principle, continuum spin systems, Kirchhoff rod.

\begin{abstract} 
We obtain the affine Euler-Poincar\'e equations by standard Lagrangian reduction and deduce the associated Clebsch-constrained variational principle. These results are illustrated in deriving the equations of motion for continuum spin systems and Kirchhoff's rod, where they provide a unified geometric interpretation.
\end{abstract}

\tableofcontents


\section{Introduction}\label{Introduction}

Lagrangian reduction by symmetry (LRS) applies when Hamilton's principle is invariant under the action of a Lie group. The implications of Lie symmetries of Hamilton's principle summons many mathematical concepts in a wealth of interesting applications of geometric mechanics in the dynamics of particles and fields, as well as in control theory. For example, the Euler-Poincar\'e (EP) reduction framework discussed in \cite{CeHoMaRa1998} and \cite{HoMaRa1998} summons all of the adjoint and coadjoint actions of Lie groups on themselves, both on their representation vector spaces and on their Lie algebras. Several applications of the EP reduction framework to classical continua are  surveyed in \cite{HoMaRa1998}. 

Continuum condensed matter theory deals with motions of a wide range of complex fluids; that is,  fluids with internal degrees of freedom, e.g., phases, spins, directors or other order parameters. Order parameters are geometrically either objects in a vector space, or coset spaces of broken Lie symmetry groups with respect to subgroups that leave these objects invariant. The evolution of the order parameters of a complex fluid is called its micro-motion. Complex fluids whose order parameters are continuous material variables are said to be perfect complex fluids ({\PCF}s). In contrast, discontinuities and singularities in the order parameter of a complex fluid are called defects and these may either be frozen into the fluid motion, or they may move relative to the material. The classical examples of complex fluids are liquid crystals, magnetic materials, and the various types of Landau theories of superfluids.  The application of the EP reduction framework to the classical {\PCF}'s is surveyed in \cite{Ho2002}. 

An interesting feature arises for the EP dynamical formulation of a perfect complex fluid whose order parameter is carried along by the flow of the material parcels. Namely, the Legendre transformation to its Hamiltonian formulation yields a noncanonical Lie-Poisson bracket that is dual to a Lie algebra possessing  generalized 2-cocycles. The generalized 2-cocycles appear in the corresponding Lie-Poisson Hamiltonian operator as covariant derivatives with respect to a Lie-algebra valued connection, in precisely the same form as found in the Yang-Mills fluid plasma for chromohydrodynamics \cite{GiHoKu1982,GiHoKu1983}, in superfluid Helium with vortices \cite{HoKu1982,HoKu1987}, in type-II superconductors \cite{HoKu1983a} and in the classical dynamics of a spin glass \cite{HoKu1988}. Following \cite{HoKu1983b} and \cite{MaWe1983}, the same Lie-Poisson Hamiltonian operator was also obtained in \cite{Ho2002} by applying Clebsch constraints to Hamilton's principle to enforce the auxiliary kinematic equations obeyed by the \PCF dynamics. While these previous approaches revealed the presence of the generalized 2-cocycles, they did not specifically reveal their source. However, recent developments in \cite{GBRa2009} provide the means of deriving these generalized 2-cocycles by showing that they are summoned when the Lie symmetry of Hamilton's principle is an affine action. Thus, affine Lie symmetry  reveals the source and guarantees the presence of generalized 2-cocycles. The present paper pursues these matters further for {\PCF}'s and for complex fluids whose defects are frozen into the flow. 

Reduction of Hamilton's principle by affine Lie symmetry is used here to construct a mathematical  framework for deriving  the affine Euler-Poincar\'e equations governing the nondissipative coupled motion and micromotion of complex fluids.
More precisely, we show how the affine Euler-Poincar\'e reduction theorem can be seen as a particular case of the general Lagrangian reduction process $TQ\rightarrow TQ/G$, that is, the case of a Lagrangian defined on a tangent bundle, and invariant under the tangent lift of a Lie group action. This is achieved by modifying the Lagrangian, in the same spirit as it is done for the Euler-Poincar\'e reduction in \cite{CeHoMaRa1998}. As a consequence, we obtain and explain the associated Clebsch-constrained variational principle in a natural way. As examples of physical interest, we apply the new variational principles to continuum spin systems and to the Kirchhoff rod.

The plan of the paper is as follows. In the first part of Section \ref{sec_AEP} we recall the relevant facts about affine Euler-Poincar\'e reduction as it applies to complex fluids and review the process of general Lagrangian reduction. Then it is shown how to obtain the affine Euler-Poincar\'e equations by standard Lagrangian reduction and the associated Clebsch-constrained variational principle is presented. In Section \ref{spin_systems}, we apply the resulting theory to the dynamics of continuum spin systems. In particular, we get the variational principle
\[
\delta\int\left( l(\nu,\gamma)+w\!\cdot\!\left(\dot\gamma-\mathbf{d}^\gamma\nu\right)\right)=0, 
\]
for variations $\delta\gamma$, $\delta w$, and constrained variations $\delta\nu=\dot\eta+[\nu,\eta]$. In Section \ref{sec_Kirchhoff}, we review some needed facts from Kirchhoff's theory of rods, using the Kirchhoff and spatial representations, and we show how the body representation can be obtained by affine Euler-Poincar\'e reduction, in the case when central potential forces are considered. Finally, by applying the theory developed in Section \ref{sec_AEP} we obtain the Clebsch-constrained variational principle for Kirchhoff's equations,
\begin{align*}
&\delta \int_{t_0}^{t_1} \left(l(\boldsymbol{\omega},\boldsymbol{\gamma},\boldsymbol{\Omega},\boldsymbol{\Gamma},\boldsymbol{\rho})
+\int_\mathcal{D} \left[ \boldsymbol{u}\!\cdot\!\left(\dot{\boldsymbol{\Omega}}-\partial_s\boldsymbol{\omega}-\boldsymbol{\Omega}\times\boldsymbol{\omega}\right) \right.\phantom{\int_{\mathcal{D}}}\right.\\
&\qquad\qquad\left.\phantom{\int_{\mathcal{D}}}\left.+\boldsymbol{w}\!\cdot\!\left(\dot{\boldsymbol{\Gamma}}+\boldsymbol{\omega}\times\boldsymbol{\Gamma}-\partial_s\boldsymbol{\gamma}-\boldsymbol{\Omega}\times\boldsymbol{\gamma}\right)+\boldsymbol{f}\!\cdot\!\left(\dot{\boldsymbol{\rho}}+\boldsymbol{\omega}\times\boldsymbol{\rho}-\boldsymbol{\gamma}\right)\right]\right)dt=0,
\end{align*}
for variations $\delta \boldsymbol{\Omega}$, $\delta\boldsymbol{\Gamma}$, $\delta \boldsymbol{\rho}$, $\delta \boldsymbol{u}$, $\delta \boldsymbol{w}$, $\delta \boldsymbol{f}$ and constrained variations 
\[
\delta\boldsymbol{\omega}=\frac{\partial\boldsymbol{\eta}}{\partial t}+\boldsymbol{\omega}\times\boldsymbol{\eta}\quad\text{and}\quad \delta\boldsymbol{\gamma}=\frac{\partial \boldsymbol{v}}{\partial t}+\boldsymbol{\omega}\times \boldsymbol{v}-\boldsymbol{\eta}\times\boldsymbol{\gamma}.
\]

\section{The affine Euler-Poincar\'e Equations via Lagrangian reduction}\label{sec_AEP}

The goal of this section is to show how the affine Euler-Poincar\'e principle can be obtained by standard Lagrange reduction of a cotangent bundle. From this alternative approach to the affine Euler-Poincar\'e equations we obtain, in a natural way, the associated Clebsch-constrained variational principle.

We begin by recalling some needed facts about affine Euler-Poincar\'e reduction for
semidirect products (see \cite{GBRa2009}). Let $V$ be a vector space and assume that the Lie group $G$ acts on the \textit{left\/} by linear maps (and hence $G$ also acts on the left on its dual space $V^*$). As a set, the semidirect product $S=G\,\circledS\,V$
is
the Cartesian product $S=G\times V$ whose group multiplication is given by
\[
(g_1,v_1)(g_2,v_2)=(g_1g_2,v_1+g_1v_2),
\]
where the action of $g\in G$ on $v\in V$ is denoted simply as $gv$.
The Lie algebra of $S$ is the semidirect product Lie algebra,
$\mathfrak{s}=\mathfrak{g}\,\circledS\,V$, whose bracket has the expression
\[
\operatorname{ad}_{(\xi_1,v_1)}(\xi_2,v_2)=[(\xi_1,v_1),(\xi_2,v_2)]=([\xi_1,\xi_2],\xi_1 v_2-\xi_2 v_1),
\]
where $\xi v$ denotes the induced action of $\mathfrak{g}$ on $V$, that is,
\[
\xi v:=\left.\frac{d}{dt}\right|_{t=0}\operatorname{exp}(t\xi)v\in
V.
\]
From the expression for the Lie bracket, it follows that for
$(\xi,v)\in\mathfrak{s}$ and $(\mu,a)\in\mathfrak{s}^*$ we have
\[
\operatorname{ad}^*_{(\xi,v)}(\mu,a)=(\operatorname{ad}^*_\xi\mu-v\diamond
a,-\xi a),
\]
where $\xi a\in V^*$ and $v\diamond a\in\mathfrak{g}^*$ are given by
\[
\xi a:=\left.\frac{d}{dt}\right|_{t=0}\operatorname{exp}(t\xi)a\quad\text{and}\quad
\langle v\diamond a,\xi\rangle_\mathfrak{g}:=-\langle \xi a,v\rangle_V,
\]
and where $\left\langle\cdot , \cdot \right\rangle_ \mathfrak{g}: \mathfrak{g}
^\ast \times \mathfrak{g}\rightarrow \mathbb{R}$ and $\left\langle \cdot ,
\cdot
\right\rangle_V: V ^\ast \times V \rightarrow \mathbb{R}$ are the duality
parings.

Given a \textit{left\/} representation of $G$ on the vector space $V^*$, we can form an affine \textit{left\/}
representation $\theta_g(a):= ga+c(g)$, where
$c\in\mathcal{F}(G,V^*)$ is a \textit{left group
one-cocycle}, that is, it verifies the property
\begin{equation}\label{cocycle_identity}
c(gh)=c(g)+gc(h)
\end{equation} 
for all $g, h \in G$. Note that
\[
\left.\frac{d}{dt}\right|_{t=0}\theta_{\operatorname{exp}(t\xi)}(a)=\xi a+\mathbf{d}c(\xi)
\]
and
\[
\langle \xi a+\mathbf{d}c(\xi),v\rangle_V=\langle \mathbf{d}c^T(v)-v\diamond
a,\xi\rangle_\mathfrak{g},
\]
where $\mathbf{d}c :\mathfrak{g}\rightarrow V^*$ is defined by
$\mathbf{d}c(\xi):=T_ec(\xi)$, and $\mathbf{d}c^T:V\rightarrow\mathfrak{g}^*$
is
defined by
\begin{equation}
\label{def_cT}
\langle \mathbf{d}c^T(v),\xi\rangle_\mathfrak{g}:=\langle
\mathbf{d}c(\xi),v\rangle_V.
\end{equation}

\paragraph{Affine Lagrangian semidirect product theory.}
\begin{itemize}
\item Assume that we have a function $L:TG\times V^*\rightarrow\mathbb{R}$
which
is \textit{left\/} $G$-invariant under the affine action $(v_h,a)\mapsto
(gv_h,\theta_g(a))=(gv_h,ga+c(g))$.
\item In particular, if $a_0\in V^*$, define the Lagrangian
$L_{a_0}:TG\rightarrow\mathbb{R}$ by 
\begin{equation}
\label{def_L_a_0}
L_{a_0}(v_g):=L(v_g,a_0).
\end{equation}
 Then $L_{a_0}$
is left invariant under the lift to $TG$ of the left action of $G_{a_0}^c$ on
$G$, where $G_{a_0}^c$ is the isotropy group of $a_0$ with respect to the
affine
action $\theta$.
\item Left $G$-invariance of $L$ permits us to define $l:\mathfrak{g}\times
V^*\rightarrow\mathbb{R}$ by
\begin{equation}
\label{def_l}
l(g^{-1}v_g,\theta_{g^{-1}}(a_0))=L(v_g,a_0).
\end{equation}
\item For a curve $g(t)\in G$, let $\xi(t):=g(t)^{-1}\dot{g}(t)$ and
define the curve $a(t)$ as the unique solution of the following affine
differential equation with time dependent coefficients
\begin{equation}\label{advection}
\dot{a}=-\xi a-\mathbf{d}c(\xi),
\end{equation}
with initial condition $a(0)=a_0$. The solution can be written as
$a(t)=\theta_{g(t)^{-1}}(a_0)$.
\end{itemize}

\begin{theorem}\label{AEPSD} With the preceding notations, the following are
equivalent:
\begin{itemize}
\item[\bf{i}] With $a_0$ held fixed, Hamilton's variational principle
\begin{equation}\label{Hamilton_principle}
\delta\int_{t_0}^{t_1}L_{a_0}(g,\dot{g})dt=0,
\end{equation}
holds, for variations $\delta g(t)$ of $g(t)$ vanishing at the endpoints.
\item[\bf{ii}] $g(t)$ satisfies the Euler-Lagrange equations for $L_{a_0}$ on
$G$.
\item[\bf{iii}] The constrained variational principle
\begin{equation}\label{Euler-Poincare_principle}
\delta\int_{t_0}^{t_1}l(\xi,a)dt=0,
\end{equation}
holds on $\mathfrak{g}\times V^*$, upon using variations of the form
\[
\delta\xi=\frac{\partial\eta}{\partial t}+[\xi,\eta],\quad \delta
a=-\eta a-\mathbf{d}c(\eta),
\]
where $\eta(t)\in\mathfrak{g}$ vanishes at the endpoints.
\item[\bf{iv}] The affine Euler-Poincar\'e equations hold on
$\mathfrak{g}\times
V^*$:
\begin{equation}\label{AEP}
\frac{\partial}{\partial t}\frac{\delta
l}{\delta\xi}=\operatorname{ad}^*_\xi\frac{\delta l}{\delta\xi}+\frac{\delta
l}{\delta a}\diamond a-\mathbf{d}c^T\left(\frac{\delta l}{\delta a}\right).
\end{equation}
\end{itemize}
\end{theorem}

See \cite{GBRa2009} for a proof and applications to complex fluids.
The affine Euler-Poincar\'e equation \eqref{AEP} can be rewritten in the form of a conservation law as
\begin{equation}\label{general_conservation_law_a_0}
\frac{\partial}{\partial t}\left[\operatorname{Ad}^*_{g^{-1}}\frac{\delta l}{\delta\xi}\right]+\mathbf{d}c^T\left(g\frac{\delta l}{\delta a}\right)=g\frac{\delta l}{\delta a}\diamond a_0.
\end{equation}

\paragraph{General Lagrangian reduction.} Consider a \textit{left\/} action $\Phi: G\times Q\rightarrow Q$ of a Lie group $G$ on a manifold $Q$. Let $L:TQ\rightarrow\mathbb{R}$ be a $G$-invariant Lagrangian under the cotangent-lifted action of $G$ on $TQ$. Because of this invariance, we get a well defined reduced Lagrangian $l : (TQ)/G \rightarrow\mathbb{R}$ satisfying
\[
l([v_q])=L(v_q).
\]
Assuming the group action is free and proper, the quotient space $(TQ)/G$ is intrinsically a vector bundle over $T(Q/G)$ with a fiber modeled on the Lie algebra $\mathfrak{g}$. Using a connection $\mathcal{A}$ on the principal bundle $\pi:Q\rightarrow Q/G$ we have a vector bundle isomorphism
\[
\alpha_\mathcal{A}:(TQ)/G\longrightarrow T(Q/G)\oplus\tilde{\mathfrak{g}},\quad [v_q]\longmapsto \alpha_\mathcal{A}([v_q]):=\left(T\pi(v_q),[q,\mathcal{A}(v_q)]_G\right)
\]
over $Q/G$, where the associated bundle $\tilde{\mathfrak{g}}:=Q\times _G\mathfrak{g}$, is defined as the quotient space of $Q\times\mathfrak{g}$ relative to the left action $(q,\xi)\mapsto (\Phi_g(q),\operatorname{Ad}_{g}\xi)$ of $G$. The elements of $\tilde{\mathfrak{g}}$ are denoted by $\bar v=[q,\xi]_G$. Using the isomorphism $\alpha_\mathcal{A}$, we can consider $l$ as a function defined on $T(Q/G)\oplus\tilde{\mathfrak{g}}$, and we write $l(x,\dot x,\bar v)$ to emphasize the dependence of $l$ on $(x,\dot x)\in T(Q/G)$ and $\bar v\in\tilde{\mathfrak{g}}$. However one should keep in mind that $x, \dot x$, and $\bar v$ cannot be considered as being independent variables unless $T (Q/G)$ and $\tilde{\mathfrak{g}}$
are trivial bundles.

We now formulate the Lagrangian reduction theorem.

\begin{theorem}\label{Lagrangian_reduction} The following conditions are equivalent:
\begin{itemize}
\item[\bf{i}] Hamilton's variational principle
\[
\delta\int_{t_0}^{t_1}L(q,\dot q)dt=0,
\]
holds, for variations $\delta q(t)$ vanishing at the endpoints.
\item[\bf{ii}] The curve $q(t)$ satisfies the Euler-Lagrange equations for $L$ on $TQ$.
\item[\bf{iii}] The reduced variational principle
\[
\delta\int_{t_0}^{t_1}l(x,\dot x,\bar v)dt=0
\]
holds, for variations $\delta x\oplus\delta^\mathcal{A} \bar v$ of the curve $x(t)\oplus \bar v(t)$, where $\delta^\mathcal{A}\bar v$ has the form
\[
\delta^\mathcal{A}\bar v=\frac{D}{Dt}\bar\eta +[\bar v,\bar \eta]+\tilde{\mathcal{B}}(\delta x,\dot x),
\]
with the boundary conditions $\delta x(t_i)=0$ and $\bar \eta(t_i)= 0$, for $i = 0, 1$.
\item[\bf{iv}] The following \textbf{vertical} and \textbf{horizontal Lagrange-Poincar\'e equations}, hold:
\begin{equation}\label{reduced_E-L}
\left\lbrace
\begin{array}{l}
\displaystyle\vspace{0.2cm}\frac{D}{Dt}\frac{\partial l}{\partial \bar v}(x,\dot x,\bar v)=\operatorname{ad}^*_{\bar v}\frac{\partial l}{\partial \bar v}(x,\dot x,\bar v)\\
\displaystyle\frac{\partial l}{\partial x}(x,\dot x,\bar v)-\frac{D}{Dt}\frac{\partial l}{\partial \dot x}(x,\dot x,\bar v)=\left\langle\frac{\partial l}{\partial \bar v}(x,\dot x,\bar v),\mathbf{i}_{\dot x}\tilde{\mathcal{B}}(x)\right\rangle.
\end{array}\right.
\end{equation}
\end{itemize}
\end{theorem}

We now comment on the various expressions appearing in parts \textbf{ii} and \textbf{iii}. In the expression
\[
\delta^\mathcal{A}\bar v=\frac{D}{Dt}\bar\eta +[\bar v,\bar \eta]+\tilde{\mathcal{B}}(\delta x,\dot x),
\]
$D/Dt$ denotes the covariant time derivative of the curve $\bar\eta(t)\in\tilde{\mathfrak{g}}$ associated to the principal connection $\mathcal{A}$, that is, for $\bar\eta(t)=[q(t),\xi(t)]_G$, we have
\[
\frac{D}{Dt}\big[q(t),\xi(t) \big]_G= \Big[ q(t),\,\dot\xi(t)-[ \mathcal{A}(q(t),\dot q(t)),\xi(t) ] \Big]_G.
\]
The bracket $[\bar v,\bar \eta]$ denotes the Lie bracket induced by $\mathfrak{g}$ on each fiber of $\tilde{\mathfrak{g}}$. The two-form $\tilde{\mathcal{B}}\in \Omega^2(Q/G,\tilde{\mathfrak{g}})$ is the curvature on the base $Q/G$ induced by the curvature form $\mathcal{B}=\mathbf{d}\mathcal{A}-[\mathcal{A},\mathcal{A}]\in\Omega^2(Q,\mathfrak{g})$ of $\mathcal{A}$. Notice that for the formulation of the Lagrange-Poincar\'e equations, the introduction of an arbitrary connection $\nabla$ on the manifold $Q/G$ is needed. For simplicity a torsion free connection is chosen. The partial derivatives
\[
\frac{\partial l}{\partial \dot x}(x,\dot x,\bar v)\in T^*_x(Q/G)\quad\text{and}\quad \frac{\partial l}{\partial \bar v}(x,\dot x,\bar v)\in \tilde{\mathfrak{g}}^*_x
\]
are the usual fiber derivatives of $l$ in the vector bundles $T(Q/G)$ and $\tilde{\mathfrak{g}}$, and
\[
\frac{\partial l}{\partial x}(x,\dot x,\bar v)\in T^*_x(Q/G)
\]
is the partial covariant derivative of $l$ relative to the given connection $\nabla$ on $Q/G$ and to the principal connection $\mathcal{A}$ on $Q$. In \eqref{reduced_E-L}, the covariant derivatives $D/Dt$ are respectively associated to the principal connection $\mathcal{A}$ and to the affine connection $\nabla$, on the dual bundles. We refer to \cite{CeMaRa2001} for details and proofs regarding the Lagrange-Poincar\'e equations.

\paragraph{Application to the affine Euler-Poincar\'e equations.} Let us assume all the hypotheses of Theorem \ref{AEPSD}. We define the Lagrangian
\begin{equation}
\label{def_bar_L}
\bar L:T(G\times V^*\times V)\rightarrow\mathbb{R},\quad \bar L(v_g,a_0,v_0,\dot a_0,\dot v_0):=L(v_g,a_0)+\langle \dot a_0,v_0\rangle,
\end{equation}
which is invariant under the tangent lift of the action
\begin{equation}\label{first_G_action}(g,a_0,v_0)\mapsto (hg,ha_0+c(h),hv_0)
\end{equation}
of $G$ on $G\times V^*\times V$, given by
\begin{equation}\label{tangent_lift_ first_G_action}
(v_g,a_0,v_0,\dot a_0,\dot v_0)\mapsto (hv_g,ha_0+c(h),hv_0,h\dot a_0,h\dot v_0).
\end{equation}
The idea in definition \eqref{def_bar_L} is to
introduce the condition that $a_0$ is conserved by making it the momentum conjugate
to a cyclic variable, just as one does for the charge in Kaluza-Klein theory.
Notice that the Euler-Lagrange equation associated to the variable $v_0$ is
\[
\dot a_0=0.
\]
Thus $a_0$ is a constant.

The variable $v_0$ is not constant, but writing the Euler-Lagrange equation for $a _0$ yields the first order equation
\begin{equation}\label{v_zero_equ}\dot v_0-\frac{\partial L}{\partial a_0}=0.
\end{equation}
Thus, the Euler-Lagrange equations for $\bar L$ are equivalent to the Euler-Lagrange equations for $L_{a_0}$ with the parameter $a_0$ fixed (together with the $v_0$-equation \eqref{v_zero_equ}).

In order to apply the Lagrangian reduction process (Theorem \ref{Lagrangian_reduction}) to the Lagrangian $\bar L$, we need to consider the principal bundle
\[
G\times V^*\times V\rightarrow (G\times V^*\times V)/G
\]
relative to the $G $-action \eqref{first_G_action}. This bundle turns out to be isomorphic to the trivial bundle
\[
G\times V^*\times V\rightarrow V^*\times V.
\]
To see this note that the diffeomorphism
\[
\psi :G\times V^*\times V\rightarrow G\times V^*\times V,\quad \psi(g,a_0,v_0):=(g,g^{-1}a_0+c(g^{-1}),g^{-1}v_0)=:(g,a,v),
\]
is $G $-equivariant relative to the action \eqref{first_G_action} and left translation of  $G $ on the first factor only, that is, 
\[
\psi(hg,ha_0+c(h),hv_0)=(hg,a,v).
\]
Using the formula
\[
T\psi^{-1}(g,\dot g,a,v,\dot a,\dot v)=(g,\dot g,ga+c(g),gv,g(\dot a+g^{-1}\dot g a+\mathbf{d}c(g^{-1}\dot g)),g(\dot v+g^{-1}\dot gv)),
\]
we obtain that the composition $L^V:=\bar L\circ T\psi^{-1}$ is given by
\begin{equation}
\label{def_L_V}
L^V(g,\dot g,a,v,\dot a,\dot v)=L(g,\dot g,ga+c(g))+\left\langle \dot a+g^{-1}\dot g a+\mathbf{d}c(g^{-1}\dot g),v\right\rangle.
\end{equation}
We shall show that the reduced Euler-Lagrange equations for the Lagrangian $L^V$ are the same as the affine Euler-Poincar\'e equations for $l$. To do this, we first calculate the reduced bundle $T(Q/G)\oplus \tilde{\mathfrak{g}}$, for $Q:=G\times V^*\times V$. Since the $G $-principal bundle $Q:=G\times V^*\times V \rightarrow  V^*\times V$ is trivial, we have $\tilde{\mathfrak{g}}=V^*\times V\times \mathfrak{g}$, and we can choose the trivial principal connection 
\begin{equation}
\label{connection_def}
\mathcal{A}(g,\dot g,a,v,\dot a,\dot v)=\dot gg^{-1}.
\end{equation}
 Thus we obtain
\[
TQ/G\cong T(Q/G)\oplus \tilde{\mathfrak{g}}\cong \mathfrak{g}\times T(V^*\times V)\ni (\xi,a,v,\dot a,\dot v),
\]
\[
[g, \dot{g}, a, v, \dot{a}, \dot{v}] \longmapsto (g ^{-1}\dot{g},a,v,\dot a,\dot v).
\]
The reduced Lagrangian induced by $L^V$ is given by
\begin{equation}
\label{def_l_V}
l^V(\xi,a,v,\dot a,\dot v)=l(\xi,a)+\langle \dot a+\xi a+\mathbf{d}c(\xi),v\rangle.
\end{equation}
Note that we can write
\begin{equation}
\label{def_l_V-sep}
l^V(\xi,a,v,\dot a,\dot v)=l(\xi,a)+\langle\theta_0(a,v),(\dot a,\dot v)\rangle+\langle\mathbf{J}(a,v),\xi\rangle,
\end{equation}
where $\mathbf{J}:V^*\times V\rightarrow\mathfrak{g}^*$ is the momentum map of the lift of the action $a\mapsto ga+c(g)$ to the cotangent bundle $T^*V^*\cong V^*\times V$, that is,
\begin{equation}
\label{momap-lift}
\mathbf{J}(a,v) = -v\diamond a+\mathbf{d}c^T(v)
\,,
\end{equation}
and $\theta_0$ is the canonical $1$-form on $T^*V^*$. Recall that $\left\langle \theta_0(a, v), (b, w) \right\rangle = \left\langle b, v \right\rangle$ for $v, w \in V $, $a, b \in V ^\ast$. 
We now compute the reduced Euler-Lagrange equations \eqref{reduced_E-L} associated to $l^V$.
The horizontal equation for $a$ is given by
\[
\frac{d}{dt}\frac{\delta l^V}{\delta \dot a}-\frac{\delta l^V}{\delta a}=0.
\]
Using the equalities
\[
\frac{\delta l^V}{\delta \dot a}=v\quad\text{and}\quad \frac{\delta l^V}{\delta a}=\frac{\delta l}{\delta  a}-\xi v,
\]
we find that this equation is equivalent to
\begin{equation}\label{Horizontal_a}
\dot v +\xi v-\frac{\delta l}{\delta a}=0.
\end{equation}
The horizontal equation for $v$ is the equation
\[
\frac{d}{dt}\frac{\delta l^V}{\delta \dot v}-\frac{\delta l^V}{\delta v}=0,
\]
which is readily seen to be equivalent to the equation
\begin{equation}\label{Horizontal_v}
\dot a+\xi a+\mathbf{d}c(\xi)=0.
\end{equation}
Since the curvature of the connection \eqref{connection_def} vanishes, the vertical equation in \eqref{reduced_E-L} becomes 
\[
\frac{d}{dt}\frac{\delta l^V}{\delta \xi}=\operatorname{ad}^*_\xi\left(\frac{\delta l^V}{\delta \xi}\right).
\]
However,  
\[
\frac{\delta l^V}{\delta \xi}=\frac{\delta l}{\delta\xi}-v\diamond a+\mathbf{d}c^T(v),
\]
and thus the vertical equation takes the form
\[
\frac{d}{dt}\frac{\delta l}{\delta \xi}-\dot v\diamond a-v\diamond \dot a+\mathbf{d}c^T(\dot v)=\operatorname{ad}^*_\xi\left(\frac{\delta l}{\delta \xi}\right)-\operatorname{ad}^*_\xi(v\diamond a-\mathbf{d}c^T(v))
\]
or, using equations \eqref{Horizontal_a} and \eqref{Horizontal_v},
\begin{align*}
\frac{d}{dt}\frac{\delta l}{\delta \xi}-\frac{\delta l}{\delta a}\diamond a+\xi v\diamond a+&v\diamond \xi a+v\diamond \mathbf{d}c(\xi)+\mathbf{d}c^T\left(\frac{\delta l}{\delta a}\right)-\mathbf{d}c^T(\xi v)\\
&=\operatorname{ad}^*_\xi\left(\frac{\delta l}{\delta \xi}\right)-\operatorname{ad}^*_\xi(v\diamond a-\mathbf{d}c^T(v)).
\end{align*}
Using the equalities
\begin{align}
\label{intermediate_identity}
\langle \xi v\diamond a+v\diamond \xi a+v\diamond \mathbf{d}c(\xi)-\mathbf{d}c^T(\xi v),\eta\rangle&=\langle (\xi\eta-\eta\xi)a,v\rangle+\langle\xi\mathbf{d}c(\eta)-\eta\mathbf{d}c(\xi),v\rangle \nonumber \\
&=\langle [\xi,\eta]a,v\rangle+\langle\mathbf{d}c([\xi,\eta]),v\rangle  
\nonumber\\
&=\langle -v\diamond a+\mathbf{d}c^T(v),[\xi,\eta]\rangle \nonumber \\
&=-\left\langle \operatorname{ad}^*_\xi\left(v\diamond a-\mathbf{d}c^T(v)\right),\eta\right\rangle,
\end{align}
we finally obtain the affine Euler-Poincar\'e equation
\[
\frac{d}{dt}\frac{\delta l}{\delta\xi}=\operatorname{ad}^*_\xi\frac{\delta l}{\delta\xi}+\frac{\delta l}{\delta a}\diamond a-\mathbf{d}c^T\left(\frac{\delta l}{\delta a}\right).
\]
The following theorem summaries the results obtained so far.

\begin{theorem}\label{lagrangian_reduction} Let $g(t)$ be a curve in $G$, fix an element $a_0\in V^*$, and define the curve $\xi(t):=g(t)^{-1}\dot g(t)$. Consider the \textit{left\/} invariant  function $L:TG\times V^*\rightarrow\mathbb{R}$ relative to the affine action
\begin{equation}\label{affine_action}
(v_g,a)\mapsto (hv_g,ha+c(h)).
\end{equation}
Define the Lagrangians 
\[
L_{a_0}:TG\rightarrow\mathbb{R},\quad\bar L:T(G\times V^*\times V)\rightarrow\mathbb{R},\quad\text{and}\quad L^V:T(G\times V^*\times V)\rightarrow\mathbb{R},
\]
$($see \eqref{def_L_a_0}, \eqref{def_bar_L}, \eqref{def_L_V}$)$ and the reduced Lagrangians
\[
l:\mathfrak{g}\times V^*\rightarrow\mathbb{R}\quad\text{and}\quad l^V:\mathfrak{g}\times T( V^*\times V)\rightarrow\mathbb{R}
\]
$($see \eqref{def_l}, \eqref{def_l_V}$)$. Let $(a(t),v(t))$ and $(a_0(t),v_0(t))$ be two curves in $V^*\times V$ related by the condition
\[
\psi(g(t),a_0(t),v_0(t))=(g(t),a(t),v(t)).
\]
For simplicity we assume that $g(0)=e$. Then the following assertions are equivalent.
\begin{itemize}
\item[\bf{i}] One of the four conditions of Theorem \ref{AEPSD} (affine Lagrangian semidirect product theory) holds, with $a(0)=a_0$.
\item[\bf{ii}] The curve $(g(t),a_0(t),v_0(t))$ is a solution of the Euler-Lagrange equations associated to $\bar L$, with $a_0(0)=a_0$ (this implies $a_0(t)=a_0$ is a constant).
\item[\bf{iii}] The curve $(g(t),a_0(t),v_0(t))$ is a critical point of the action
\[
\int_{t_0}^{t_1}\bar L(g,\dot{g}, a_0,v_a,\dot a_0,\dot v_0)dt
\]
for variations $\delta g(t),\delta a_0(t)$, and $\delta v_0(t)$ vanishing at the endpoints.
\item[\bf{iv}] The curve $(g(t),a(t),v(t))$ is a solution of the Euler-Lagrange equations associated to $L^V$, with $a(0)=a_0$.
\item[\bf{v}]  The curve $(g(t),a(t),v(t))$ is a critical point of the action
\[
\int_{t_0}^{t_1}L^V(g,\dot{g},a,v,\dot a,\dot v)dt
\]
for variations $\delta g(t),\delta a(t)$, and $\delta v(t)$ vanishing at the endpoints.
\item[\bf{vi}] The curve $(\xi(t),a(t),v(t))$ is solution of the Lagrange-Poincar\'e (or reduced Euler-Lagrange) equations \eqref{reduced_E-L} associated to $l^V$, with $a(0)=a_0$. In our case, these equations read
\begin{equation}
\label{LP_equ_specific}
\left\{
\begin{array}{l}
\vspace{0.2cm}\displaystyle\frac{d}{dt}\frac{\delta l^V}{\delta \xi}=\operatorname{ad}^*_\xi\left(\frac{\delta l^V}{\delta \xi}\right)\\
\displaystyle\frac{d}{dt}\frac{\delta l^V}{\delta \dot a}-\frac{\delta l^V}{\delta a}=0,\quad \displaystyle\frac{d}{dt}\frac{\delta l^V}{\delta \dot v}-\frac{\delta l^V}{\delta v}=0.
\end{array}\right.
\end{equation}

\item[\bf{vii}]  The curve $(\xi(t),a(t),v(t))$ is a critical point of the action
\[
\int_{t_0}^{t_1} l^V(\xi,a,v,\dot a,\dot v)dt
\]
for variations $\delta a(t)$, and $\delta v(t)$ vanishing at the endpoints, and variations $\delta\xi$ of the form
\[
\delta \xi=\frac{\partial \eta}{\partial t}+[\xi,\eta],
\]
where $\eta(t)\in\mathfrak{g}$ vanishes at the endpoints.
\end{itemize}
\end{theorem}
\noindent \textbf{Proof.} This theorem is a direct consequence of the discussion above and of the Lagrangian reduction theorem (Theorem \ref{Lagrangian_reduction}). To see this, it suffices to show that the constrained variational principle in \textbf{vii} coincides with the one in Theorem \ref{Lagrangian_reduction} \textbf{iii}  for the particular case when $Q=G\times V^*\times V$. Indeed, we can identify the equivalence class $[(g,a,v),\xi]_G\in\tilde{\mathfrak{g}}=(G\times V^*\times V)\times_G\mathfrak{g}$ with the element $(a,v,\operatorname{Ad}_{g^{-1}}\xi)\in V^*\times V\times\mathfrak{g}$. Using the trivial connection $\mathcal{A}(g,\dot g,a,v,\dot a,\dot v)=\dot gg^{-1}$, the associated covariant derivative in $\tilde{\mathfrak{g}}\cong V^*\times V\times\mathfrak{g}$ is readily seen to be given by
\[
\frac{D}{Dt}(a(t),v(t),\xi(t))=(a(t),v(t),\dot\xi(t)).
\]
Thus, since the curvature of $\mathcal{A}$ vanishes, the general expression $\delta^\mathcal{A}\bar v=\frac{D}{Dt}\bar\eta+[\bar v,\bar\eta]+\tilde{\mathcal{B}}(\delta x,\dot x)$ reads simply
\[
\delta\xi=\frac{\partial \eta}{\partial t}+[\xi,\eta].
\]
This proves the result.$\qquad\blacksquare$

\begin{remark} \normalfont
Note that the curves $v (t)$, $v _0 (t)$ do not appear in point \textbf{i}. However, we can always recover them. The equation for $v_0(t) $ is given by \eqref{v_zero_equ}. The evolution for $v(t) $ is implicitly contained in the definition of $\psi$, namely, $v(t) = g (t) ^{-1}v_0(t) $, where $v_0(t) $ is the solution of \eqref{v_zero_equ} and $g(t)$ is the $G $-part of the base integral curve for the Lagrangians $L_{a_0} $, $L^V $, or $\overline{L}$ (they all give the same $g(t)$ by construction). Equivalently, the equation for $v(t) $ can be obtained by writing the Euler-Lagrange equation for $L^V$ relative to the variable $a $. An easy computation yields
\begin{equation}\label{v_equ}
\dot{v} - g ^{-1}\frac{\partial L }{\partial a _0} + (g ^{-1}\dot{g})v = 0.
\end{equation}
\end{remark}

We now check directly the equivalence between parts \textbf{vi} and \textbf{vii}. If
\[
S:=\int_{t_0}^{t_1}l^V(\xi,a,\dot a,v,\dot v)dt=\int_{t_0}^{t_1}\left( l(\xi,a)+\langle v,\dot a+\xi a+\mathbf{d}c(\xi)\rangle \right)dt
\]
we get 
\begin{align*}
\delta S&=\int_{t_0}^{t_1}\left( \left\langle\frac{\delta l}{\delta\xi},\delta\xi\right\rangle+\left\langle\frac{\delta l}{\delta a},\delta a\right\rangle+\langle\delta v,\dot a+\xi a+\mathbf{d}c(\xi)\rangle  \right. \\
& \left. \phantom{\left\langle\frac{\delta l}{\delta\xi},\delta\xi\right\rangle}
\qquad  +\langle v,\delta\dot a \rangle+\langle v,(\delta \xi)a+\xi(\delta a)+\mathbf{d}c(\delta\xi)\rangle \right)dt\\
&=\int_{t_0}^{t_1}\left( \left\langle\frac{\delta l}{\delta\xi},\delta\xi\right\rangle+\left\langle\frac{\delta l}{\delta a},\delta a\right\rangle+\langle\delta v,\dot a+\xi a+\mathbf{d}c(\xi)\rangle \right. \\
& \left. \phantom{\left\langle\frac{\delta l}{\delta\xi},\delta\xi\right\rangle}
\qquad  -\langle \dot v,\delta a \rangle-\langle \xi v,\delta a\rangle-\langle v\diamond a,\delta\xi\rangle+\langle\mathbf{d}c^T(v),\delta\xi\rangle \right) dt\\
&=\int_{t_0}^{t_1}\left( \left\langle-\frac{d}{dt}\frac{\delta l}{\delta\xi}+\operatorname{ad}^*_\xi\frac{\delta l}{\delta \xi}+\frac{d}{dt}(v\diamond a-\mathbf{d}c^T(v))-\operatorname{ad}^*_\xi(v\diamond a-\mathbf{d}c^T(v)),\eta\right\rangle+ \right. \\
& \left. \phantom{\left\langle\frac{\delta l}{\delta\xi},\delta\xi\right\rangle}
\qquad  +\langle\delta v,\dot a+\xi a+\mathbf{d}c(\xi)\rangle+\left\langle \frac{\delta l}{\delta a}-\dot v-\xi v,\delta a\right\rangle \right) dt,
\end{align*}
where, in the last equality, we have used the relation $\delta\xi=\dot\eta-[\xi,\eta]$. Thus, $\delta S=0$ is equivalent to
\[
\dot a+\xi a+\mathbf{d}c(\xi)=0,\quad \frac{\delta l}{\delta a}-\dot v-\xi v=0,
\]
and
\[
\frac{d}{dt}\frac{\delta l}{\delta\xi}=\operatorname{ad}^*_\xi\frac{\delta l}{\delta \xi}+\frac{d}{dt}(v\diamond a-\mathbf{d}c^T(v))-\operatorname{ad}^*_\xi(v\diamond a-\mathbf{d}c^T(v)).
\]
Using \eqref{intermediate_identity} and the the first two equations, the preceding equality reads
\[
\frac{d}{dt}\frac{\delta l}{\delta \xi}=\operatorname{ad}^*_\xi\frac{\delta l}{\delta\xi}+\frac{\delta l}{\delta a}\diamond a-\mathbf{d}c^T\left(\frac{\delta l}{\delta a}\right).
\]
Thus, the variational principle in \textbf{vii} holds if and only if the following system 
\begin{eqnarray}
\left\lbrace
\begin{array}{l}
\displaystyle\vspace{0.2cm}\frac{d}{dt}\frac{\delta l}{\delta \xi}=\operatorname{ad}^*_\xi\frac{\delta l}{\delta\xi}+\frac{\delta l}{\delta a}\diamond a-\mathbf{d}c^T\left(\frac{\delta l}{\delta a}\right)\\
\displaystyle\dot a+\xi a+\mathbf{d}c(\xi)=0,\quad \frac{\delta l}{\delta a}-\dot v-\xi v=0
\end{array}\right.
\label{LP-eqns}
\end{eqnarray}
is satisfied. These equations are readily seen to be the Lagrange-Poincar\'e equations associated to $l^V$, and are equivalent to the affine Euler-Poincar\'e equations (\ref{AEP}), together with the advection equation for the variable $a$, and the equation for $v$.
 
\begin{remark} \normalfont
Recall the formula
\[
\frac{ \delta l^V}{\delta \xi} = \frac{\delta l}{\delta\xi}-v\diamond a+\mathbf{d}c^T(v) = \frac{\delta l}{\delta\xi} + \mathbf{J}(a, v).
\]
If the horizontal equations for $a$ and $v$ hold (the second line in the system \eqref{LP-eqns}),  a direct computation shows that we have the equivalences
\begin{align*}
\frac{\delta l^V}{\delta\xi}=constant 
&\quad\Longleftrightarrow \quad \frac{d}{dt}\frac{\delta l}{\delta\xi}=\operatorname{ad}^*_\xi\frac{\delta l}{\delta \xi}+\frac{\delta l}{\delta a}\diamond a-\mathbf{d}c^T\left(\frac{\delta l}{\delta a}\right) \\
&\quad\Longleftrightarrow \quad \frac{d}{dt} \frac{\delta l^V}{\delta\xi} = \operatorname{ad}_ \xi ^\ast  \frac{\delta l^V}{\delta \xi} \\
&\quad\Longleftrightarrow \quad \frac{\delta l^V}{\delta\xi}(t) = \operatorname{Ad}^\ast _{g(t)} \left[\frac{\delta l^V}{\delta\xi} \right]_{t=0},
\end{align*}
where $g(t) $ is determined by $g(t) ^{-1} \dot{g}(t) = \xi(t) $, $g(0)=e$.
The last equivalence is a direct consequence of the general formula
\[
\frac{d}{dt} \Big(\operatorname{Ad}^\ast_{h(t) ^{-1}} \mu(t) \Big) 
= \operatorname{Ad}^\ast_{h(t) ^{-1}} \left( \frac{d}{dt} \mu(t) - \operatorname{ad}^\ast_{ \xi(t)} \mu(t) \right), \quad \xi(t) = h (t) ^{-1} \dot{h} (t),
\]
where $t \mapsto h(t)$ and $t \mapsto \mu(t)$ are smooth curves in $G$  and $\mathfrak{g}^\ast$, respectively.

Let $\rho\in \mathfrak{g}^\ast$ and $b \in V ^\ast$ be given. Assume that the linear system $w \diamond b - \mathbf{d}c ^T(w) = \rho$ has always a solution. Under these circumstances we can always choose the initial condition $v(0) \in V$ such that the constant in $\delta l^V/ \delta \xi = constant$ vanishes. 
This is a convenient initial condition because it is preserved by the flow. One can think of this initial condition giving rise to $\delta l^V/ \delta \xi = 0$ geometrically as one does with the vanishing angular momentum condition in the falling cat problem. 
\end{remark}

\section{Continuum spin system}\label{spin_systems}

In this section we apply the previous theory to the dynamics of spin systems.

The order parameter for a spin system is any Lie algebra, such as $\mathfrak{o}=\mathfrak{su}(2)$. Thus, besides the usual continuum degrees of freedom, Hamilton's principle for a spin system will depend on smooth functions of space and time that take values in a Lie algebra $\mathfrak{o}$. Obvious examples are the magnetization vectors of magnetized fluids and spin-glass fluids, and a case can be made for also regarding $He^3$-$A$ and $He^3$-$B$ as spin systems. Another example, would be a fluid of oriented nano-particles, whose order parameter would take values on the unit sphere. Yet another example would comprise a fluid system of particles, each of which is a 2-level qubit. The order-parameter dependence of the spin-system's prescribed interaction energy determines its Hamilton's principle.

It is shown in \cite{GBRa2009} that the equations of motion, as well as their variational and Hamiltonian structures, can be obtained by affine Euler-Poincar\'e or Lie-Poisson reduction. Given a manifold $\mathcal{D}$ and a Lie group $\mathcal{O}$, we consider the Fr\'echet Lie group $\mathcal{F}(\mathcal{D},\mathcal{O})$ of smooth functions $\chi:\mathcal{D}\rightarrow\mathcal{O}$, and the vector space $V^*=\Omega^1(\mathcal{D},\mathfrak{o})$ of one-forms with values in the Lie algebra $\mathfrak{o}$ of $\mathcal{O}$. When $\mathcal{D}$ is endowed with a volume form $\mu$, this space is in a natural way the dual of the space $V=\mathfrak{X}(\mathcal{D},\mathfrak{o}^*)$ of $\mathfrak{o}^*$-valued vector fields on $\mathcal{D}$, the duality pairing being given by contraction and integration over $\mathcal{D}$.  Let $\chi\in G$ act on $\gamma\in V^*$ via the \textit{left\/} affine representation 
\[
\gamma\mapsto \operatorname{Ad}_{\chi}\gamma+\chi\mathbf{d}\chi^{-1}.
\]
This action is natural in the sense that it coincides with the usual gauge transformation of a connection on the trivial principal bundle $\mathcal{O}\times\mathcal{D}$. The Lagrangian of a spin system is a function
\[
L_{\gamma_0}:TG\rightarrow\mathbb{R},
\]
such that $L(\nu_\psi,\gamma):=L_\gamma(\nu_\psi)$ is invariant under the left affine action of $\chi\in G$ on $TG\times V^*$ given by
\[
(\nu_\psi,\gamma)\mapsto \left(\chi\nu_\psi,\operatorname{Ad}_{\chi}\gamma+\chi \mathbf{d}\chi^{-1}\right).
\]
We can now apply the general theory of affine Euler-Poincar\'e reduction to the group $G=\mathcal{F}(\mathcal{D},\mathcal{O})$, the space $V^*=\Omega^1(\mathcal{D},\mathfrak{o})$, and the group one-cocycle $c\in\mathcal{F}(G,V^*)$ given by
\[
c(\chi)=\chi\mathbf{d}\chi^{-1}.
\]
One can check that the cocycle property
\[
c(\chi\psi)=c(\chi)+\operatorname{Ad}_{\chi}c(\psi)
\]
holds, relative to the representation $\gamma\mapsto \operatorname{Ad}_\psi\gamma$ of $G$ on $V^*$. As in the abstract theory, we use the notation $\theta_\chi\gamma=\operatorname{Ad}_\chi\gamma+\chi\mathbf{d}\chi^{-1}$ for the affine representation.

For $w\in V$ and $\gamma\in V^*$, the associated diamond operation is given by
\[
w\diamond\gamma=\operatorname{Tr}(\operatorname{ad}^*_\gamma w)=\operatorname{ad}^*_{\gamma_i} w^i.
\]
Since $c(\chi)=\chi\mathbf{d}\chi^{-1}$ for $\nu\in\mathfrak{g}$ and $w\in V$, from (\ref{def_cT}) we obtain the relations
\[
\mathbf{d}c(\nu)=-\mathbf{d}\nu\quad\text{and}\quad\mathbf{d}c^T(w)=\operatorname{div}w.
\]
Thus, the expression for the cotangent-lift momentum map (\ref{momap-lift}),
\[
\mathbf{d}c^T\left(\frac{\delta l}{\delta\gamma}\right)
-
\frac{\delta l}{\delta\gamma}\diamond\gamma
=
\mathbf{J}\left(\gamma,\frac{\delta l}{\delta\gamma}\right) 
,
\]
appearing with the opposite sign in the right hand side of the affine Euler-Poincar\'e equations (\ref{AEP}), in this case reads
\begin{equation}
\label{momap1}
\operatorname{div}\frac{\delta l}{\delta\gamma}
-
\operatorname{Tr}\left(\operatorname{ad}^*_\gamma \frac{\delta l}{\delta\gamma}\right)
=
\mathbf{J}\left(\gamma,\frac{\delta l}{\delta\gamma}\right) .
\end{equation}
Remarkably, the cotangent-lift momentum map in this case recovers an expression known as the \textit{covariant divergence} associated to $\gamma$, defined on $w\in \mathfrak{X}(\mathcal{D},\mathfrak{o}^*)$ by
\begin{equation}\label{cov_div}
\operatorname{div}^\gamma w=\operatorname{div}w-\operatorname{Tr}\left(\operatorname{ad}^*_\gamma w\right).
\end{equation}
Making use of these observations, the affine Euler-Poincar\'e
equations \eqref{AEP} become
\begin{equation}\label{AEP_spin_chain}
\frac{\partial}{\partial t}\frac{\delta
l}{\delta\nu}-\operatorname{ad}^*_\nu\frac{\delta
l}{\delta\nu}+\operatorname{div}^\gamma\frac{\delta l}{\delta\gamma}=0.
\end{equation}
Since the infinitesimal action of $\nu$ on $\gamma$ induced by the representation $\gamma\mapsto\operatorname{Ad}_{\chi}\gamma$ is
\[
\gamma\nu=\operatorname{ad}_\nu\gamma,
\]
the advection equation \eqref{advection} for $\gamma$ is
\begin{equation}\label{gamma-advection_spin_chain}
\frac{\partial}{\partial t}\gamma-\mathbf{d}\nu+\operatorname{ad}_\nu\gamma=0.
\end{equation}
Here also, the expressions associated to the affine representation combine in a nice way, since the quantity $(\mathbf{d}\nu-\operatorname{ad}_\nu\gamma)$ is known as the \textit{covariant derivative} of $\nu$ associated to $\gamma$, and is usually denoted by $\mathbf{d}^\gamma\nu$. The $\gamma$-advection equation (\ref{gamma-advection_spin_chain}) can thus be rewritten as
\begin{equation}\label{advection_spin_chain}
\frac{\partial}{\partial t}\gamma=\mathbf{d}^\gamma\nu.
\end{equation}
In the particular case $\mathcal{D}=\mathbb{R}^3$ and $\mathcal{O}=SO(3)$, equations \eqref{AEP_spin_chain} and \eqref{advection_spin_chain} appear in the context of the \textit{macroscopic description of spin glasses}, see equations (28) and (29) in \cite{Dz1980} and references therein. See also equations (3.9), (3.10) in \cite{IsKoPe1994}, system (1) in \cite{Iv2000} and references therein for an application to \textit{magnetic media}. In this context, the quantity $\delta l/\delta\nu$ is interpreted as the \textit{spin density}, $\nu$ is the \textit{infinitesimal spin rotation} and the curvature $B:=\mathbf{d}^\gamma\gamma$ is the \textit{disclination density}.

\begin{remark} \normalfont
Thus, the covariant derivatives appearing in the Hamiltonian operator as generalized 2-cocycles in the dynamics of spin systems arise directly from the affine Lie symmetry of the Hamilton's principle.
\end{remark}  

\paragraph{Variational principles for spin systems.} We now specialize to the case of spin systems some results obtained abstractly in Theorem \ref{lagrangian_reduction}. Start with a Lagrangian 
\[
L_{\gamma_0}:T\mathcal{F}(\mathcal{D},\mathcal{O})\rightarrow\mathbb{R},
\]
describing the dynamics of a spin system and depending on a parameter $\gamma_0\in V^*=\Omega^1(\mathcal{D},\mathfrak{o})$. Suppose that the function $L$, defined by $L(\nu_\psi,\gamma_0):=L_{\gamma_0}(\nu_\psi)$, is invariant under the left affine action of $\chi\in \mathcal{F}(\mathcal{D},\mathcal{O})$ on $T\mathcal{F}(\mathcal{D},\mathcal{O})\times \Omega^1(\mathcal{D},\mathfrak{o})$ given by
\[
(\nu_\psi,\gamma)\mapsto \left(\chi\nu_\psi,\operatorname{Ad}_\chi\gamma+\chi\mathbf{d}\chi^{-1}\right).
\]
We will also use the notation $(\psi,\dot\psi)$ instead of $\nu_\psi$. The Lagrangians $\overline{L}$ and $L^V$ (see \eqref{def_bar_L}, \eqref{def_L_V}) defined on $T\left(\mathcal{F}(\mathcal{D},\mathcal{O})\times\Omega^1(\mathcal{D},\mathfrak{o})\times\mathfrak{X}(\mathcal{D},\mathfrak{o}^*)\right)$ are given by
\[
\overline{L}(\psi,\dot\psi,\gamma_0,\dot \gamma_0,w_0,\dot w_0)=L(\psi,\dot\psi,\gamma_0)+\int_\mathcal{D}\left(\dot\gamma_0\!\cdot w_0\right)\mu
\]
and
\[
L^V(\psi,\dot\psi,\gamma,\dot \gamma,w,\dot w)=L(\psi,\dot\psi,\theta_{\psi}\gamma)+\int_\mathcal{D}w\!\cdot\!\left(\dot\gamma-\mathbf{d}^\gamma\psi^{-1}\dot\psi\right)\mu.
\]
The reduced Lagrangian $l^V$ (see \eqref{def_l_V}) defined on $\mathcal{F}(\mathcal{D},\mathfrak{o})\times T(\Omega^1(\mathcal{D},\mathfrak{o})\times\mathfrak{X}(\mathcal{D},\mathfrak{o}^*))$ is
\[
l^V(\nu,\gamma,w,\dot\gamma,\dot w)=l(\nu,\gamma)+\int_\mathcal{D}w\!\cdot\!\left(\dot\gamma-\mathbf{d}^\gamma\nu\right)\mu.
\]
We are now ready to state several variational formulations of the dynamics of spin systems obtained from Theorem \ref{Lagrangian_reduction}.
\medskip

Let $\chi(t)$ be a curve in the group $G=\mathcal{F}(\mathcal{D},\mathcal{O})$ and fix an element $\gamma_0\in V^*=\Omega^1(\mathcal{D},\mathfrak{o})$. The choice $\gamma_0=0$ is allowed and important. Define the curve $\nu(t):=\chi(t)^{-1}\dot\chi(t)\in\mathcal{F}(\mathcal{D},\mathfrak{o})$. Let $(\gamma_0(t),w_0(t))$ and $(\gamma(t),w(t))$ be two curves in $\Omega^1(\mathcal{D},\mathfrak{o})\times\mathfrak{X}(\mathcal{D},\mathfrak{o}^*)$ related by the conditions
\[
\operatorname{Ad}_{\chi(t)^{-1}}\gamma_0(t)+\chi(t)^{-1}\mathbf{d}\chi(t)=\gamma(t)\quad\text{and}\quad \operatorname{Ad}^*_{\chi(t)}w_0(t)=w(t),
\]
and such that $\gamma_0(0)=\gamma_0$. Suppose, for simplicity, that $\chi(0)=e$.
Then the following are equivalent:
\begin{itemize}
\item[\bf{i}] Hamilton's variational principle
\[
\delta \int_{t_0}^{t_1} L_{\gamma_0}(\chi,\dot\chi)dt=0,
\]
holds, for variations $\delta\chi(t)$ vanishing at the endpoints.
\item[\bf{ii}] Hamilton's variational principle
\[
\delta \int_{t_0}^{t_1} \left(L(\chi,\dot\chi,\gamma_0)+\int_\mathcal{D}\dot\gamma_0\!\cdot w_0\right)dt=0,
\]
holds, for variations $\delta\chi(t), \delta\gamma_0(t)$, and $\delta w_0(t)$ vanishing at the endpoints.
\item[\bf{iii}] Hamilton's variational principle
\[
\delta \int_{t_0}^{t_1} \left(L\left(\chi,\dot\chi,\theta_{\chi}\gamma\right)+\int_\mathcal{D}w\!\cdot\!\left(\dot\gamma-\mathbf{d}^{\gamma}\chi^{-1}\dot\chi\right)\right)dt=0,
\]
holds, for variations $\delta\chi(t), \delta\gamma(t)$, and $\delta w(t)$ vanishing at the endpoints.
\item[\bf{iv}] The reduced variational principle
\[
\delta \int_{t_0}^{t_1} \left(l(\nu,\gamma)+\int_\mathcal{D}w\!\cdot\!\left(\dot\gamma-\mathbf{d}^{\gamma}\nu\right)\right)dt=0,
\]
holds, for variations $\delta\gamma(t)$, and $\delta w(t)$ vanishing at the endpoints, and variations $\delta\nu(t)$ of the form
\[
\delta\nu=\frac{\partial\eta}{\partial t}+[\nu,\eta],
\]
where $\eta(t)$ vanishes at the endpoints.
\item[\bf{v}] The reduced variational principle
\[
\delta \int_{t_0}^{t_1}l(\nu,\gamma)dt=0,
\]
holds, for variations $\delta\nu(t)$ and $\delta\gamma(t)$ of the form
\[
\delta\nu=\frac{\partial\eta}{\partial t}+[\nu,\eta]\quad\text{and}\quad\delta\gamma=\mathbf{d}^\gamma\eta
\]
where $\eta(t)$ vanishes at the endpoints.
\end{itemize}

Note that the curve $\gamma(t)$ is not present in part \textbf{i}, but it can be recovered from the curve $\chi(t)$, through the relation
\begin{equation}
\label{gamma_evolution}
\gamma(t)=\theta_{\chi(t)^{-1}}\gamma_0=\operatorname{Ad}_{\chi(t)^{-1}}\gamma_0+\chi(t)^{-1}\mathbf{d}\chi(t).
\end{equation}
The curve $\gamma_0(t)=\gamma_0$ is a constant. The curve $w(t)$ is not present in the parts $\textbf{i}$ and $\textbf{v}$, but it can be recovered by solving the equation
\[
\dot w-\operatorname{ad}^*_\nu w-\frac{\delta l}{\delta \gamma}=0,
\]
obtained from stationarity under variations with respect to $\gamma$. The variable $w$ is the momentum  canonically conjugate to the 1-form $\gamma$, just as the electric field is canonically conjugate to the vector potential in electromagnetism and its Yang-Mills generalization for any gauge group. The $w$-equation may be written equivalently as
\[
\dot w-\operatorname{ad}^*_{\chi^{-1}\dot\chi}w
-\operatorname{Ad}^*_{\chi}\frac{\partial L}{\partial\gamma_0}
=0\,.
\]
The curve $w_0(t)$ is related to $w(t)$ through the relation
\[
w_0(t)=\operatorname{Ad}^*_{\chi(t)^{-1}}w(t),
\]
or it can be obtained directly by solving the equation
\[
\dot w_0-\frac{\partial L}{\partial\gamma_0}=0.
\]
All these variational principles are equivalent to the affine Euler-Poincar\'e equation \eqref{AEP_spin_chain} together with the equations for $\gamma$ and $w$.

The conservation law \eqref{general_conservation_law_a_0} for spin reads
\begin{equation}
\label{conservation_law_spin}
\frac{\partial}{\partial t}\left(\operatorname{Ad}^*_{\chi^{-1}}\frac{\delta l}{\delta\nu}\right)+\operatorname{div}^{\gamma_0}\left(\chi\frac{\delta l}{\delta\gamma}\right)=0,
\end{equation}
where $\gamma_0$ is the initial value of $\gamma$. In the important case when $\gamma_0=0$, one finds the conservation law,
\[
\frac{\partial}{\partial t}\left(\operatorname{Ad}^*_{\chi^{-1}}\frac{\delta l}{\delta\nu}\right)+\operatorname{div}\left(\chi\frac{\delta l}{\delta\gamma}\right)=0,
\]
cf. \S 10 in \cite{Ho2008}, especially equation (10.1.6).

\begin{remark} \label{remark_sec3}
\normalfont
The curvature of $\gamma(t)$ corresponds to the density of defects in the order parameter. If initially $\gamma_0 = 0$, then the curvature of $\gamma(t) $ remains zero for all $t$ (see \eqref{gamma_evolution}) and no defects arise. However, a non-zero initial value for the connection $\gamma_0 \neq 0$ allows nonzero curvature, corresponding to defects, which in turn alter the spin dynamics of the system \eqref{conservation_law_spin}, now rewritten equivalently as,
\begin{equation}
\frac{\partial}{\partial t}\left(\operatorname{Ad}^*_{\chi^{-1}}\frac{\delta l}{\delta\nu}\right)+\operatorname{div}\left(\chi\frac{\delta l}{\delta\gamma}\right)= \operatorname{Tr} \left(\operatorname{ad}^\ast_{ \gamma_0} \left(\chi \frac{\delta l}{ \delta \gamma} \right) \right).
\label{spin-defect-coupling}
\end{equation}
The right hand side of this equation shows the effect of a non-zero initial value of the connection $\gamma$ on the dynamics of the spin density. 
\end{remark}

\begin{remark} \label{remark1_sec3}
\normalfont
Using the advection equation \eqref{gamma-advection_spin_chain}, that is, $\frac{ \partial \gamma}{\partial t}=\mathbf{d}\nu+[\gamma,\nu]$, if $c $ is a closed loop in $\mathcal{D}$, we get
\[
\frac{d}{dt}\oint_c\gamma=\oint_c(\mathbf{d}\nu-\operatorname{ad}_\nu\gamma)=-\oint_c\operatorname{ad}_\nu\gamma.
\]
In particular, if $\mathcal{O}$ is Abelian, we get
\[
\frac{d}{dt}\oint_c\gamma=0.
\]

Returning to a general Lie group $\mathcal{O}$, we calculate the dynamical equation for the defect density $B:=\mathbf{d}^\gamma\gamma=\mathbf{d}\gamma+[\gamma,\gamma]$ by taking the $\gamma$-covariant exterior derivative (covariant curl). We get
\begin{align*}
\dot B&=\mathbf{d}\dot\gamma+[\dot\gamma,\gamma]+[\gamma,\dot\gamma]=\mathbf{d}\left(\mathbf{d}\nu+[\gamma,\nu]\right)+\left[\mathbf{d}\nu+[\gamma,\nu],\gamma\right]+\left[\gamma,\mathbf{d}\nu+[\gamma,\nu]\right]\\
&=[\mathbf{d}\gamma+[\gamma,\gamma],\nu]=[B,\nu].
\end{align*}
Thus the advection equation for the curvature reads
\[
\frac{\partial B}{ \partial t}+\operatorname{ad}_\nu B=0
\]
which implies
\begin{equation}
\frac{d}{dt} \iint_S B = - \iint \operatorname{ad}_{ \nu} B, 
\label{defect-spin-coupling}
\end{equation}
for any compact two-dimensional submanifold $S $ of $\mathcal{D}$.

In particular, if $\mathcal{O}$ is Abelian, the advection equation for the curvature becomes the conservation law $\partial B/ \partial t = 0 $. For 
example,  if $\operatorname{dim}\mathcal{D}=3$ and $\mathcal{O}=S^1$, we can define the vector field $\mathbf{B}=(\star B)^\sharp$  and the integral of the left hand side of the previous equation is the flux of $\mathbf{B}$; we get in this case
\[
\frac{d}{dt}\iint_S(\mathbf{B}\!\cdot\!\mathbf{n})dS=0.
\]

For a general Lie group, equations \eqref{spin-defect-coupling} and  \eqref{defect-spin-coupling} govern the nonlinear coupling between spin and defect density dynamics. If initially $\gamma_0 = 0$ then \eqref{gamma_evolution} shows that  the curvature $B(t)$ of $\gamma(t) $ remains zero for all $t$ and no defects arise. However, a non-zero initial value for the connection $\gamma_0 \neq 0$ allows nonzero curvature, corresponding to the initial presence of defects. The defects alter the spin dynamics of the system in \eqref{spin-defect-coupling}, which in turn feed back nonlinearly to influence the defect  dynamics in  (\ref{defect-spin-coupling}). The right hand side of equation (\ref{defect-spin-coupling}) shows the non-commutative effect of the spin density on the defect density due to a non-zero initial value $\gamma_0 \neq 0$ of the connection $\gamma$. Namely, non-commutativity $ \iint \operatorname{ad}_{ \nu} B\ne0$ generates defects when  $\gamma_0 \neq 0$. 

\end{remark}

\section{Elastic filament dynamics and Kirchhoff's theory}\label{sec_Kirchhoff}

The equations of motion for continuum mechanical systems can be formulated in three representations: material or Lagrangian, spatial or Eulerian, and convective or body. For filaments, there is a fourth representation, due to Kirchhoff, that is often very convenient.

\subsection{Kirchhoff's theory and spatial representation}

In this subsection we briefly recall the relevant aspects of Kirchhoff's theory of rods (see \cite{DiLiMa1996} for a modern formulation). We shall apply it here to models that depend explicitly on order-parameter variables. In \cite{ElGBHoPuRa2009} one can also deal with models that have nonlocal interactions by using affine Euler-Poincar\'e reduction instead of Kirchhoff's theory.

The Kirchhoff rod is described by a vector $\boldsymbol{r}(s)$ and an orthonormal basis of director $\{\mathbf{d}_1(s),\mathbf{d}_2(s),\mathbf{d}_3(s)\}$. The curve $\boldsymbol{r}(s)$ is interpreted as the configuration of the center line, the triad $\{\mathbf{d}_k\}$ can be interpreted as providing information concerning the orientation of the material cross-section of the rod. Given a fixed orthonormal basis $\{\mathbf{E}_1,\mathbf{E}_2,\mathbf{E}_3\}$, we define the transformation matrix $\Lambda(s)\in SO(3)$ by requiring the equality
\begin{equation}\label{def_basis_d}
\mathbf{d}_i(s)=\Lambda^k_i(s)\mathbf{E}_k,\;\;i=1,2,3.
\end{equation}
From the variables $\boldsymbol{r}$ and $\Lambda$ we define the \textit{body angular velocity} $\boldsymbol{\omega}$ and the \textit{linear velocity} $\boldsymbol{\gamma}$ by
\begin{equation}\label{body_quantities}
\widehat{\boldsymbol{\omega}}=\Lambda^{-1}\dot{\Lambda}\;\;\text{and}\;\;\boldsymbol{\gamma}=\Lambda^{-1}\dot{\boldsymbol{r}},
\end{equation}
where $\widehat{\phantom{\Omega}}: (\mathbb{R}^3, \times ) \to (\mathfrak{so}(3), [\,,])$ is the Lie algebra isomorphism given by $\widehat{\boldsymbol{u}}\boldsymbol{v}=\boldsymbol{u} \times \boldsymbol{v}$ for all $\boldsymbol{v}\in\mathbb{R}^3$.
Thus, in an orthonormal basis of $\mathbb{R}^3$ and $\boldsymbol{u} \in \mathbb{R}^3$, the $3\times3$ antisymmetric matrix $u:=\widehat{\boldsymbol{u}} \in \mathfrak{so}(3)$ has entries
\begin{equation}
u_{jk}
=
(\widehat{\boldsymbol{u}})_{jk}
=
-\epsilon_{jkl}\boldsymbol{u}^l
\, .
\label{hatmap-components}
\end{equation}
Note that the relations \eqref{body_quantities} can be rewritten using the group structure of $SE(3)$ as
\[
(\widehat{\boldsymbol{\omega}},\boldsymbol{\gamma})=(\Lambda,\boldsymbol{r})^{-1}(\dot{\Lambda},\dot{\boldsymbol{r}}).
\]
The linear momentum density $\mathbf{p}$ is defined as $\mathbf{p}(s)=\rho_d(s) \boldsymbol{\gamma}(s)$, where $\rho_d(s)$ is the local mass density of the rod. In that case, the kinetic energy due to linear motion $K_{lin}$ is given by
\[ 
K_{lin}= \frac{1}{2} \int \rho_d(s) \|\dot{\boldsymbol{r}}(s) \|^2 \mbox{d} s 
= 
\frac{1}{2} \int \rho_d(s) \|  \Lambda^{-1} \dot{\boldsymbol{r}}(s)\|^2 \mbox{d} s
=  \frac{1}{2} \int \rho_d(s) \|\boldsymbol{\gamma}(s) \|^2 \mbox{d} s. 
\] 
Consequently, we have
\[
\mathbf{p}=\frac{\delta K_{lin}}{\delta \boldsymbol{\gamma}}.
\]
The local angular momentum in the \textit{body} frame $\{ \mathbf{d}_1(s), \mathbf{d}_2(s), \mathbf{d}_3(s) \}$ is defined by $\boldsymbol{\pi}(s):=\mathbb{I}(s)\boldsymbol{\omega}(s)$, where $\mathbb{I}(s)$ is the local value of the inertia tensor expressed in body coordinates. Relative to body coordinates we have 
$\boldsymbol{\pi}^i(s):=\mathbb{I}^i_{j}(s) \boldsymbol{\omega}^j(s)$. Thus the local kinetic energy due to rotation is given by 
\[ 
K_{rot}= \frac{1}{2} \int \boldsymbol{\omega}(s)\!\cdot\!\mathbb{I}(s)\boldsymbol{\omega} \mbox{d} s 
\]
and hence
\[ 
\boldsymbol{\pi} = \mathbb{I} \boldsymbol{\omega} = \frac{\delta K_{rot}}{\delta \boldsymbol{\omega} }. 
\]
In Kirchhoff's approach, the conservation laws are written in terms of variables expressed in the fixed \textit{spatial} frame $\{ \mathbf{E}_1, \mathbf{E}_2, \mathbf{E}_3\}$. To distinguish it from $\mathbf{p}$ and $\boldsymbol{\pi}$ which were expressed in the \textit{body} frame 
$\{ \mathbf{d}_1(s), \mathbf{d}_2(s), \mathbf{d}_3(s) \}$, we shall denote the 
same vectors in the fixed spatial frame $\{ \mathbf{E}_1, \mathbf{E}_2, \mathbf{E}_3\}$ by $\mathbf{p}^{(\mathbf{E})}$ and $\boldsymbol{\pi}^{(\mathbf{E})}$. Thus, \eqref{def_basis_d} yields the relations
\[
\boldsymbol{\pi}=\boldsymbol{\pi}^i\mathbf{d}_i=\boldsymbol{\pi}^{( \mathbf{E}),k}\mathbf{E}_k\;\;\text{and}\;\;\mathbf{p}=\mathbf{p}^i\mathbf{d}_i=\mathbf{p}^{( \mathbf{E}),k}\mathbf{E}_k,
\]
where 
\begin{equation} 
\boldsymbol{\pi}^{(\mathbf{E}),k}=\Lambda^k_i\mathbb{I}^i_{j} \boldsymbol{\omega}^j = \left[ \Lambda  \mathbb{I} \boldsymbol{\omega}\right]^k =  \left[ \Lambda  \frac{\delta K_{rot}}{\delta \boldsymbol{\omega}}  \right]^k\;\;\text{and}\;\;\mathbf{p}^{( \mathbf{E}),k}=\rho_d\Lambda^k_i\boldsymbol{\gamma}^i=\left[\Lambda  \frac{\delta K_{lin}}{\delta \boldsymbol{\gamma}}  \right]^k.
\label{piomconnect0}
\end{equation} 
Thus, the vectors $\mathbf{p}^{(\mathbf{E})}$ and $\boldsymbol{\pi}^{( \mathbf{E})}$  of \textit{body} linear and angular momenta expressed in the \textit{spatial} frame are connected to the local body quantities as 
\begin{equation} 
\boldsymbol{\pi}^{(\mathbf{E})}= \Lambda  \frac{\delta K_{rot}}{\delta \boldsymbol{\omega}}\;\;\text{and}\;\;\mathbf{p}^{(\mathbf{E})}=\Lambda\frac{\delta K_{lin}}{\delta \boldsymbol{\gamma}}.
\label{piomconnect}
\end{equation} 

In general, it is assumed for physical reasons, that the Lagrangian in Kirchhoff's formulation has the form 
\begin{equation} 
l(\boldsymbol{\omega},\boldsymbol{\gamma},\boldsymbol{\Omega},\boldsymbol{\Gamma})=K_{lin}(\boldsymbol{\gamma})+K_{rot}(\boldsymbol{\omega})-E(\boldsymbol{\Omega},\boldsymbol{\Gamma}) \, , 
\label{lkirchhoff}
\end{equation} 
where the potential energy $E(\boldsymbol{\Omega}, \boldsymbol{\Gamma})$
is usually taken to be a quadratic function of the deformations $\Omega:=\Lambda^{-1}\partial_s\Lambda$ and $\boldsymbol{\Gamma}:=\Lambda^{-1}\partial_s\boldsymbol{r}$, but more complex expressions are possible as well; we shall not restrict the functional form of that dependence. In this case, the body  
forces $\mathbf{n}={\delta l}/{\delta \boldsymbol{\Gamma}}$  and torques $\mathbf{m}={\delta l}/{\delta \boldsymbol{\Omega}}$ are connected to the transformed quantities $\mathbf{n}^{(\mathbf{E})}, \mathbf{m}^{(\mathbf{E})}$ in Kirchhoff's theory as 
\begin{equation} 
\mathbf{n}^{(\mathbf{E})}=\Lambda \frac{\delta l}{\delta \boldsymbol{\Gamma}}, 
\quad \mathbf{m}^{(\mathbf{E})}=\Lambda \frac{\delta l}{\delta \boldsymbol{\Omega}}.
\label{forcetorque} 
\end{equation}
The balances of linear and angular momenta in Kirchhoff's approach (cf. eqs. (2.3.5) and (2.3.6) of \cite{DiLiMa1996}) are given by
\begin{align}
&\frac{\partial \mathbf{p}^{(\mathbf{E})}}{\partial t}  + \frac{\partial \mathbf{n}^{(\mathbf{E})}}{\partial s} =\mathbf{f},
\label{linmom}
\\
&\frac{\partial \boldsymbol{\pi}^{(\mathbf{E})}}{\partial t} + \frac{\partial  \mathbf{m}^{(\mathbf{E})}}{\partial s}+
\frac{\partial \boldsymbol{r}}{\partial s} \times \mathbf{n}^{(\mathbf{E})}
=\mathbf{l}.
\label{angmom}
\end{align}
where $\mathbf{f}$ and $\mathbf{l}$ denote the \textit{external forces} and \textit{torques}, respectively.

We now rewrite equations \eqref{linmom}, \eqref{angmom} in terms of spatial variables, using the formula 
\begin{equation} 
{\rm Ad}^*_{(\Lambda,\boldsymbol{r})^{-1}} \left( \boldsymbol{\mu}, \boldsymbol{\eta} \right) 
= \left( \Lambda \boldsymbol{\mu}+ \boldsymbol{r} \times \Lambda \boldsymbol{\eta}, 
\Lambda \boldsymbol{\eta} \right). 
\end{equation}
for the coadjoint action of $SE(3)$ on $\mathfrak{se}(3)^*$. The spatial momenta -- denoted by a superscript $(S)$ -- become 
\begin{align} 
\left(\boldsymbol{\pi}^{(S)},\mathbf{p}^{(S)}\right):&=\operatorname{Ad}^*_{(\Lambda,\boldsymbol{r})^{-1}}\left(\frac{\delta l}{\delta \boldsymbol{\omega}},\frac{\delta l}{\delta \boldsymbol{\gamma}}\right)=\left( 
\Lambda \frac{\delta l}{\delta \boldsymbol{\omega}}+ \boldsymbol{r} \times \Lambda \frac{\delta l}{\delta \boldsymbol{\gamma}},\Lambda \frac{\delta l}{\delta \boldsymbol{\gamma}}
\right)\nonumber\\
&=\left( \boldsymbol{\pi}^{(\mathbf{E})}+ \boldsymbol{r} \times \mathbf{p}^{(\mathbf{E})},\mathbf{p}^{(\mathbf{E})} 
\right), 
\label{spacemomenta} 
\end{align} 
and the spatial torques  $\mathbf{m}^{(S)}$ and forces $\mathbf{n}^{(S)}$  are  
\begin{align} 
\left(\mathbf{m}^{(S)},\mathbf{n}^{(S)}\right)
:&=\operatorname{Ad}^*_{(\Lambda,\boldsymbol{r})^{-1}}
\left(\frac{\delta l}{\delta \boldsymbol{\Omega}},\frac{\delta l}{\delta \boldsymbol{\Gamma}}\right)
=\left( 
\Lambda \frac{\delta l}{\delta \boldsymbol{\Omega}}+ \boldsymbol{r} \times \Lambda \frac{\delta l}{\delta \boldsymbol{\Gamma}},\Lambda \frac{\delta l}{\delta \boldsymbol{\Gamma}}
\right)\nonumber 
\\
&=\left( \mathbf{m}^{(\mathbf{E})}+ \boldsymbol{r} \times \mathbf{n}^{(\mathbf{E})},\mathbf{n}^{(\mathbf{E})} 
\right). 
\label{spaceforces} 
\end{align} 
The conservation laws in Kirchhoff theory may now be written as 
\begin{equation}\label{Kirchcons0}
\frac{\partial}{\partial t} \left(\mathbf{\boldsymbol{\pi}}^{(S)},\mathbf{p}^{(S)}\right)
+
\frac{\partial}{\partial s}\left(\mathbf{m}^{(S)},\mathbf{n}^{(S)}\right)
=(\mathbf{T}, \mathbf{f}),  
\end{equation} 
where $\mathbf{T}:=\boldsymbol{r}\times\mathbf{f}+\mathbf{l}$.

\subsection{Affine Euler-Poincar\'e formulation}

We now show that, in the case of \textit{potential} forces, we can obtain these equations by affine Euler-Poincar\'e reduction.

Consider the group $G=\mathcal{F}([0,L],SE(3))$ of all smooth maps from the interval $[0,L]$ into the group $SE(3)$, whose elements are denoted by $(\Lambda,\boldsymbol{r})$. The space of affine advected quantities is chosen to be the dual vector space
\[
V^*:=\Omega^1([0,L],\mathfrak{se}(3))\oplus\mathcal{F}([0,L],\mathbb{R}^3)\ni (\boldsymbol{\Omega},\boldsymbol{\Gamma},\boldsymbol{\rho}).
\]
Note that since $[0,L]$ is one dimensional, the space $\Omega^1([0,L],\mathfrak{se}(3))$ can be identified with the space $\mathcal{F}([0,L],\mathfrak{se}(3))$ of smooth maps with values in $\mathfrak{se}(3)$. We choose to think of these functions as one-forms, since it is that interpretation that generalizes to the higher dimensional case.
We consider the affine representation $\theta$ of the group $G$ on $V^*$, given by
\[
\theta_{(\Lambda,\boldsymbol{r})}(\boldsymbol{\Omega},\boldsymbol{\Gamma},\boldsymbol{\rho})=(\Lambda,\boldsymbol{r})(\boldsymbol{\Omega},\boldsymbol{\Gamma},\boldsymbol{\rho})+c(\Lambda,\boldsymbol{r}),
\]
where the first term denotes the representation of $G$ on $V^*$ defined by
\begin{align}\label{repres_molecular_strand}
(\Lambda,r)(\boldsymbol{\Omega},\boldsymbol{\Gamma},\boldsymbol{\rho})&=\left(\operatorname{Ad}_{(\Lambda,\boldsymbol{r})}(\boldsymbol{\Omega},\boldsymbol{\Gamma}),\Lambda\boldsymbol{\rho}\right)
\end{align}
and the second term is the \textit{group one-cocycle} given by
\begin{equation}\label{cocycle_Kirchhoff}
c(\Lambda,\boldsymbol{r})=\left((\Lambda,\boldsymbol{r})\partial_s(\Lambda,\boldsymbol{r})^{-1},-\boldsymbol{r}\right),
\end{equation}
where $s\in [0,L]$. The fact that $c$ is a group one-cocycle relative to the representation \eqref{repres_molecular_strand} (that is, it satisfies \eqref{cocycle_identity}) follows from a direct computation. Note that the first two components of $c$ form the left version of the cocycle appearing in the theory of complex fluids, see \cite{GBRa2009}. The infinitesimal linear action of the Lie algebra element $(\boldsymbol{\omega},\boldsymbol{\gamma})\in\mathcal{F}([0,L],\mathfrak{se}(3))$ is
\[
(\boldsymbol{\omega},\boldsymbol{\gamma})(\boldsymbol{\Omega},\boldsymbol{\Gamma},\boldsymbol{\rho})=(\operatorname{ad}_{(\boldsymbol{\omega},\boldsymbol{\gamma})}(\boldsymbol{\Omega},\boldsymbol{\Gamma}),\boldsymbol{\omega}\boldsymbol{\rho})=(\boldsymbol{\omega}\times \boldsymbol{\Omega},\boldsymbol{\omega}\times\boldsymbol{\Gamma}-\boldsymbol{\Omega}\times\boldsymbol{\gamma},\boldsymbol{\omega}\times\boldsymbol{\rho})
\]
and the associated diamond operation is
\[
(\boldsymbol{u},\boldsymbol{w},\boldsymbol{f})\diamond (\boldsymbol{\Omega},\boldsymbol{\Gamma},\boldsymbol{\rho})=(\boldsymbol{u}\times\boldsymbol{\Omega}+\boldsymbol{w}\times\boldsymbol{\Gamma}+\boldsymbol{f}\times\boldsymbol{\rho},\boldsymbol{w}\times\boldsymbol{\Omega}).
\]
Concerning the cocycle $c$, we have the formulas
\[
\mathbf{d}c(\boldsymbol{\omega},\boldsymbol{\gamma})=(-\partial_s\boldsymbol{\omega},-\partial_s\boldsymbol{\gamma},-\boldsymbol{\gamma}),\;\;\mathbf{d}c^T(\boldsymbol{u},\boldsymbol{w},\boldsymbol{f})=(\partial_s\boldsymbol{u},\partial_s\boldsymbol{w}-\boldsymbol{f}).
\]

Using all these expressions, the affine Euler-Poincar\'e equations \eqref{AEP} become in this case
\begin{equation}\label{AEP_Kirchhoff}
\left\{
\begin{array}{l}
\vspace{0.2cm}\displaystyle\left(\partial_t + \boldsymbol{\omega}\times\right)\frac{\delta l}{\delta \boldsymbol{\omega}} + \left(\partial_s + \boldsymbol{\Omega}\times\right)\frac{\delta l}{\delta \boldsymbol{\Omega}}= \frac{\delta l}{\delta \boldsymbol{\gamma}}\times\boldsymbol{\gamma} + \frac{\delta l}{\delta\boldsymbol{\Gamma}}\times\boldsymbol{\Gamma} + \frac{\delta l}{\delta \boldsymbol{\rho}}\times\boldsymbol{\rho}\\
\displaystyle\left(\partial_t + \boldsymbol{\omega}\times\right)\frac{\delta l}{\delta\boldsymbol{\gamma}}+\left( \partial_s + \boldsymbol{\Omega}\times\right)\frac{\delta l}{\delta \boldsymbol{\Gamma}}=\frac{\delta l}{\delta \boldsymbol{\rho}}.
\end{array}\right.
\end{equation}
The advection equation \eqref{advection} for Kirchhoff's theory are
\begin{equation}\label{advection_Kirchhoff}
\left\lbrace
\begin{array}{l}
\displaystyle\vspace{0.2cm}\dot{\boldsymbol{\Omega}}+\boldsymbol{\omega}\times\boldsymbol{\Omega}=\partial_s\boldsymbol{\omega}\\
\displaystyle\vspace{0.2cm}\dot{\boldsymbol{\Gamma}}+\boldsymbol{\omega}\times\boldsymbol{\Gamma}=\partial_s\boldsymbol{\gamma}+\boldsymbol{\Omega}\times\boldsymbol{\gamma}\\
\displaystyle\vspace{0.2cm}\dot{\boldsymbol{\rho}}+\boldsymbol{\omega}\times\boldsymbol{\rho}=\boldsymbol{\gamma}.
\end{array}\right.
\end{equation}

If we fix the initial values $\boldsymbol{\Omega}_0$, $\boldsymbol{\Gamma}_0$, $\boldsymbol{\rho}_0$, the conservation law \eqref{general_conservation_law_a_0} reads
\[
\frac{\partial}{\partial t}\left[\operatorname{Ad}^*_{(\Lambda,\boldsymbol{r})^{-1}}\left(\frac{\delta l}{\delta\boldsymbol{\omega}},\frac{\delta l}{\delta\boldsymbol{\gamma}}\right)\right]+\operatorname{div}^{(\boldsymbol{\Omega}_0,\boldsymbol{\Gamma}_0)}\left[\operatorname{Ad}^*_{(\Lambda,\boldsymbol{r})^{-1}}\left(\frac{\delta l}{\delta\boldsymbol{\Omega}},\frac{\delta l}{\delta\boldsymbol{\Gamma}}\right)\right]=\left(\Lambda\frac{\delta l}{\delta\boldsymbol{\rho}}\diamond\boldsymbol{\rho}_0,\Lambda\frac{\delta l}{\delta\boldsymbol{\rho}}\right)
\]
or, more explicitly,
\begin{align}
\label{general_conservation_law_Kirchhoff}
&\frac{\partial}{\partial t}\left[\operatorname{Ad}^*_{(\Lambda,\boldsymbol{r})^{-1}}\left(\frac{\delta l}{\delta\boldsymbol{\omega}},\frac{\delta l}{\delta\boldsymbol{\gamma}}\right)\right]+\frac{\partial}{\partial s}\left[\operatorname{Ad}^*_{(\Lambda,\boldsymbol{r})^{-1}}\left(\frac{\delta l}{\delta\boldsymbol{\Omega}},\frac{\delta l}{\delta\boldsymbol{\Gamma}}\right)\right] \nonumber \\
&\quad =
\left(\left(\Lambda\frac{\delta l}{\delta \boldsymbol{\Omega}}+\boldsymbol{r}\times \Lambda\frac{\delta l}{\delta \boldsymbol{\Gamma}}\right)\times\boldsymbol{\Omega}_0
+\Lambda\frac{\delta l}{\delta \boldsymbol{\Gamma}}\times\boldsymbol{\Gamma}_0
+\Lambda\frac{\delta l}{\delta \boldsymbol{\rho}}\times\boldsymbol{\rho}_0,
\Lambda\frac{\delta l}{\delta \boldsymbol{\Gamma}}\times\boldsymbol{\Gamma}_0
+\Lambda\frac{\delta l}{\delta \boldsymbol{\rho}}
\right) \nonumber \\
& \quad = \left(\mathbf{m}^{(S)} \times \boldsymbol{\Omega}_0
+ \mathbf{n}^{(S)} \times \boldsymbol{\Gamma}_0
+ \Lambda\frac{\delta l}{\delta \boldsymbol{\rho}}\times\boldsymbol{\rho}_0,
\mathbf{n}^{(S)} \times \boldsymbol{\Gamma}_0 + \Lambda\frac{\delta l}{\delta \boldsymbol{\rho}} \right) = :(\mathbf{T}, \mathbf{f})
\end{align}
using \eqref{spaceforces}. Note that the right hand side of this equation is formed by an external torque $\mathbf{T}$ and an external force $\mathbf{f}$ that recover \eqref{Kirchcons0} in terms of the spatial variables
\[
\left(\boldsymbol{\pi}^{(S)},\mathbf{p}^{(S)}\right)=\operatorname{Ad}^*_{(\Lambda,\boldsymbol{r})^{-1}}\left(\frac{\delta l}{\delta\boldsymbol{\omega}},\frac{\delta l}{\delta\boldsymbol{\gamma}}\right)\quad\text{and}\quad \left(\mathbf{m}^{(S)},\mathbf{n}^{(S)}\right)=\operatorname{Ad}^*_{(\Lambda,\boldsymbol{r})^{-1}}\left(\frac{\delta l}{\delta\boldsymbol{\Omega}},\frac{\delta l}{\delta\boldsymbol{\Gamma}}\right).
\]

If $\boldsymbol{\Omega}_0 = 0$, $\boldsymbol{\Gamma}_0 = 0$, $\boldsymbol{\rho}_0 = 0$, the conservation law \eqref{general_conservation_law_Kirchhoff} simplifies to 
\begin{equation}
\label{special_Kirchhoff_conservation_law}
\frac{\partial}{\partial t}\left[\operatorname{Ad}^*_{(\Lambda,\boldsymbol{r})^{-1}}\left(\frac{\delta l}{\delta\boldsymbol{\omega}},\frac{\delta l}{\delta\boldsymbol{\gamma}}\right)\right]+\frac{\partial}{\partial s}\left[\operatorname{Ad}^*_{(\Lambda,\boldsymbol{r})^{-1}}\left(\frac{\delta l}{\delta\boldsymbol{\Omega}},\frac{\delta l}{\delta\boldsymbol{\Gamma}}\right)\right]=\left(0,\Lambda\frac{\delta l}{\delta\boldsymbol{\rho}}\right).
\end{equation}
Note that in this case, there is no external torque.  Conversely, non-zero initial values $\boldsymbol{\Omega}_0$, $\boldsymbol{\Gamma}_0$, $\boldsymbol{\rho}_0$ produce the external torque in \eqref{general_conservation_law_Kirchhoff}. This torque phenomenon for the Kirchhoff rod  corresponds to the effect of non-zero curvature (defect density) in Remark \ref{remark_sec3} for spin systems. The corresponding connection for the Kirchhoff rod is given by the pair $(\boldsymbol{\Omega}, \boldsymbol{\Gamma})$. 

When the derivative $\delta l/\delta\boldsymbol{\gamma}$ is proportional to $\boldsymbol{\gamma}$, which is the case for Kirchhoff's Lagrangian, equations \eqref{special_Kirchhoff_conservation_law} can be rewritten as
\begin{align}
&\frac{\partial \mathbf{p}^{(\mathbf{E})}}{\partial t}  + \frac{\partial \mathbf{n}^{(\mathbf{E})}}{\partial s} =\Lambda\frac{\delta l}{\delta\boldsymbol{\rho}},
\\
&\frac{\partial \boldsymbol{\pi}^{(\mathbf{E})}}{\partial t} + \frac{\partial  \mathbf{m}^{(\mathbf{E})}}{\partial s}+
\frac{\partial \boldsymbol{r}}{\partial s} \times \mathbf{n}^{(\mathbf{E})}
=\Lambda\frac{\delta l}{\delta\boldsymbol{\rho}}\times\boldsymbol{r},
\end{align}
and we recover the balances of linear and angular momenta in Kirchhoff's approach when central potential forces are considered; that is, when the potential depends only on the magnitude of distance $|\boldsymbol{r}|=|\boldsymbol{\rho}|$. The variable $\boldsymbol{\rho}=\Lambda^{-1}\boldsymbol{r}$ is the vector $\boldsymbol{r}$ from the center of coordinates to a point on the rod, as seen from an orthogonal frame fixed at that point on the moving rod.

\medskip
Returning to the general case, recall that the reduced Lagrangian is of the form
\begin{equation}\label{Lagrangian_Kirchhoff_rod}
l(\boldsymbol{\omega},\boldsymbol{\gamma},\boldsymbol{\Omega},\boldsymbol{\Gamma},\boldsymbol{\rho})=K_{rot}(\boldsymbol{\omega})+K_{lin}(\boldsymbol{\gamma})+E(\boldsymbol{\Omega},\boldsymbol{\Gamma},\boldsymbol{\rho}).
\end{equation}
Since $|\boldsymbol{r}|=|\boldsymbol{\rho}|$, one can replace a central force in $\boldsymbol{r}$ by one in $\boldsymbol{\rho}$. For example, one could discuss the dynamics of a Kirchhoff rod moving in a gravitational field, such as a tether hanging from an orbiting satellite and feeling the force of Earth's gravity.

\paragraph{Generalizations.} In fact, our approach generalizes, without additional difficulties, to the case when the variables $\Lambda$ and $r$ are defined on an arbitrary manifold $\mathcal{D}$ and take values in an arbitrary Lie group $\mathcal{O}$ and an arbitrary vector space $E$, respectively. Correspondingly, the other variables are
\[
\Omega\in\Omega^1(\mathcal{D},\mathfrak{o}),\;\;\Gamma\in \Omega^1(\mathcal{D},E),\;\;\rho\in\mathcal{F}(\mathcal{D},E),
\]
where $\mathfrak{o}$ denotes the Lie algebra of $\mathcal{O}$. The symmetry group $G=\mathcal{F}([0,L],SE(3))$ of Kirchhoff's theory generalizes to the group $G=\mathcal{F}(\mathcal{D},S)$, where $S$ is the semidirect product $S=\mathcal{O}\,\circledS\,E$ relative to a \textit{left\/} representation of $\mathcal{O}$ on $E$. Thus, the group multiplication is given by
\[
(\chi,v)(\psi,w)=\left(\chi\psi,v+\chi w\right).
\]
The space of affine advected quantities is chosen to be the vector space $V^*:=\Omega^1(\mathcal{D},\mathfrak{s})\oplus\mathcal{F}(\mathcal{D},E)\ni (\Omega,\Gamma,\rho)$, dual to the space $V=\mathfrak{X}(\mathcal{D},\mathfrak{s}^*)\oplus\mathcal{F}(\mathcal{D},E^*)\ni (m,n,f)$. The affine representation $\theta$ of the group $G$ on $V^*$ has the same expression as before, namely, it is given by
\[
\theta_{(\Lambda,r)}(\Omega,\Gamma,\rho)=(\Lambda,r)(\Omega,\Gamma,\rho)+c(\Lambda,r),
\]
where the first term denotes the representation of $G$ on $V^*$ defined by
\[
(\Lambda,r)(\Omega,\Gamma,\rho)=\left(\operatorname{Ad}_{(\Lambda,r)}(\Omega,\Gamma),\Lambda\rho\right)=\left(\operatorname{Ad}_{\Lambda}\Omega,\Lambda\Gamma-(\operatorname{Ad}_{\Lambda}\Omega)r,\Lambda\rho\right)
\]
and the second term is the generalization of the cocycle \eqref{cocycle_Kirchhoff} given by
\[
c(\Lambda,r)=\left((\Lambda,r)\mathbf{d}(\Lambda,r)^{-1},-r\right)=(\Lambda\mathbf{d}\Lambda^{-1},-\mathbf{d}r-(\Lambda\mathbf{d}\Lambda^{-1})r,-r).
\]
Using the formulas
\[
(u,w,f)\diamond (\Omega,\Gamma,\rho)=(\operatorname{ad}^*_{\Omega_i}u^i+w^i\diamond\Gamma_i+f\diamond \rho,-\Omega_iw^i)
\]
and
\[
\mathbf{d}c(\omega,\gamma)=(-\mathbf{d}\omega,-\mathbf{d}\gamma,-\gamma),\quad 
\mathbf{d}c^T(u,w,f)=(\operatorname{div}(u),\operatorname{div}(w)-f),
\]
the affine Euler-Poincar\'e equations read
\begin{equation}\label{AEP_Kirchhoff_generalized}
\left\lbrace\begin{array}{l}
\displaystyle\vspace{0.2cm}\left(\frac{\partial}{\partial t}-\operatorname{ad}^*_{\omega}\right)\frac{\delta l}{\delta\omega}+\operatorname{div}^\Omega\frac{\delta l}{\delta\Omega}=\frac{\delta l}{\delta\gamma}\diamond\gamma+\frac{\delta l}{\delta\Gamma^i}\diamond\Gamma^i+\frac{\delta l}{\delta\rho}\diamond\rho\\
\displaystyle\left(\frac{\partial}{\partial t}+\omega\right)\frac{\delta l}{\delta\gamma}+\operatorname{div}^\Omega\frac{\delta l}{\delta\Gamma}=\frac{\delta l}{\delta\rho},
\end{array}\right.
\end{equation}
where $\operatorname{div}^\Omega$ is the covariant divergence defined in \eqref{cov_div}. The advection equations \eqref{advection} become in our case
\begin{equation}
\left\lbrace\begin{array}{l}
\displaystyle\vspace{0.2cm}\dot\Omega=\mathbf{d}^\Omega\omega\,,\\
\displaystyle\vspace{0.2cm}\dot\Gamma+\omega\Gamma=\mathbf{d}^\Omega\gamma\,,\\
\displaystyle\vspace{0.2cm}\dot\rho+\omega\rho=\gamma\,,
\end{array}\right.
\end{equation}
where
\[
\mathbf{d}^\Omega\omega=\mathbf{d}\omega+\operatorname{ad}_\Omega\omega\quad\text{and}\quad\mathbf{d}^\Omega\gamma=\mathbf{d}\gamma+\Omega\gamma.
\]
When specialized to the case $\mathcal{D}=[0,L]$ and $S=SE(3)$, these equations recover \eqref{AEP_Kirchhoff} and \eqref{advection_Kirchhoff} of Kirchhoff's theory.

\subsection{Variational principles for the Kirchhoff rod}

In this paragraph, we specialize some results obtained abstractly in Theorem \ref{lagrangian_reduction} to the case of the Kirchhoff rod. In order to give a more transparent vision of the underlying
geometric structures, we consider the generalization described
above, that is, we replace the interval $[0, L]$ by an arbitrary manifold $\mathcal{D}$ and we replace $SE(3)$ by the semidirect product $S = \mathcal{O}\,\circledS\,E$ of a Lie group $\mathcal{O}$ with a
left representation space $E$.

Start with a Lagrangian
\[
L_{(\Omega_0,\Gamma_0,\rho_0)}:T\mathcal{F}(\mathcal{D},S)\rightarrow\mathbb{R},
\]
depending on the parameters
\[
(\Omega_0,\Gamma_0,\rho_0)\in \Omega^1(\mathcal{D},\mathfrak{s})\oplus \mathcal{F}(\mathcal{D},E).
\]
Suppose that the function $L$, defined by
\[
L(\Lambda,\dot\Lambda,r,\dot r,\Omega_0,\Gamma_0,\rho_0):=L_{(\Omega_0,\Gamma_0,\rho_0)}(\Lambda,\dot\Lambda,r,\dot r),
\]
is invariant under the left affine action of $(\chi,v)\in \mathcal{F}(\mathcal{D},S)$ on
\[
(\Lambda,\dot\Lambda,r,\dot r,\Omega,\Gamma,\rho)\in T\mathcal{F}(\mathcal{D},S)\times\left(\Omega^1(\mathcal{D},\mathfrak{s})\oplus \mathcal{F}(\mathcal{D},E)\right)
\]
given by
\[
\left[\begin{array}{l}\Lambda\\ \dot\Lambda\\ r\\ \dot r\\ \Omega\\ \Gamma\\ \rho
\end{array}\right]\longmapsto 
\left[\begin{array}{l}\chi\Lambda\\ \chi\dot\Lambda\\v+\chi r \\ \chi\dot r\\ \operatorname{Ad}_\chi\Omega+\chi \mathbf{d}\chi^{-1}\\ \chi\Gamma-(\operatorname{Ad}_\chi\Omega)v-\mathbf{d}^{\chi\mathbf{d}\chi^{-1}}v\\ \chi\rho-v
\end{array}\right].
\]
This seemingly complicated action is nothing else than the affine action \eqref{affine_action} written for our $G: = \mathcal{F}( \mathcal{D}, S)$ and $V^* := \Omega^1(\mathcal{D},\mathfrak{s})\oplus \mathcal{F}(\mathcal{D},E)$.
Such a function $L$ is completely determined by the expression of the reduced Lagrangian $l$, for example the Lagrangian \eqref{Lagrangian_Kirchhoff_rod} of Kirchhoff's rod.

The Lagrangians $\overline{L}$ and $L^V$ defined on the tangent bundle
\[
T\left[\mathcal{F}(\mathcal{D},S)\times \left(\Omega^1(\mathcal{D},\mathfrak{s})\oplus\mathcal{F}(\mathcal{D},E)\right)\times \left(\mathfrak{X}(\mathcal{D},\mathfrak{s}^*)\oplus\mathcal{F}(\mathcal{D},E^*)\phantom{{\!\!}^1}\right) \right]
\]
are given by (see \eqref{def_bar_L}, \eqref{def_L_V})
\begin{align*}
&\overline{L}(\Lambda,\dot\Lambda,r,\dot r,\Omega_0,\Gamma_0,\rho_0,\dot \Omega_0,\dot\Gamma_0,\dot\rho_0,u_0,w_0,f_0,\dot u_0,\dot w_0,\dot f_0)\\
&\quad\quad =L(\Lambda,\dot\Lambda,r,\dot r,\Omega_0,\Gamma_0,\rho_0)+\int_\mathcal{D}\dot\Omega_0\cdot u_0+\dot\Gamma_0\!\cdot\! w_0+\dot\rho_0\!\cdot\!f_0
\end{align*}
and
\begin{align*}
L^V(\Lambda,&\dot\Lambda,r,\dot r,\Omega,\Gamma,\rho,\dot \Omega,\dot\Gamma,\dot\rho,u,w,f,\dot u,\dot w,\dot f)\\
=&L(\Lambda,\dot\Lambda,r,\dot r,\operatorname{Ad}_\Lambda\Omega+\Lambda\mathbf{d}\Lambda^{-1},\Lambda\Gamma-(\operatorname{Ad}_\Lambda\Omega)r-\mathbf{d}^{\Lambda\mathbf{d}\Lambda^{-1}}r,\Lambda\rho-r)\\
&\quad+\int_\mathcal{D}u\!\cdot\!\left(\dot\Omega-\mathbf{d}^\Omega\omega\right)+w\!\cdot\!\left(\dot\Gamma+\omega\Gamma-\mathbf{d}^\Omega\gamma\right)+f\!\cdot\!\left(\dot\rho+\omega\rho-\gamma\right),
\end{align*}
where $\omega=\Lambda^{-1}\dot\Lambda$ and $\gamma=\Lambda^{-1}\dot r$. The reduced Lagrangian $l^V$ defined on
\[
\mathcal{F}(\mathcal{D},\mathfrak{s})\times T\left[\left(\Omega^1(\mathcal{D},\mathfrak{s})\oplus\mathcal{F}(\mathcal{D},E)\right)\times \left(\mathfrak{X}(\mathcal{D},\mathfrak{s}^*)\oplus\mathcal{F}(\mathcal{D},E^*)\phantom{\!\!^1}\right)\right]
\]
is (see \eqref{def_l_V})
\begin{align*}
l^V(\omega,&\gamma,\Omega,\Gamma,\rho,\dot \Omega,\dot\Gamma,\dot\rho,u,w,f,\dot u,\dot w,\dot f)\\
=&l(\omega,\gamma,\Omega,\Gamma,\rho)+\int_\mathcal{D}u\!\cdot\!\left(\dot\Omega-\mathbf{d}^\Omega\omega\right)+w\!\cdot\!\left(\dot\Gamma+\omega\Gamma-\mathbf{d}^\Omega\gamma\right)+f\!\cdot\!\left(\dot\rho+\omega\rho-\gamma\right).
\end{align*}
We are now ready to state the various variational formulations of the dynamics of Kirchhoff's rod obtained from Theorem \ref{lagrangian_reduction}.

Let $(\Lambda(t),r(t))$ be a curve in the group $\mathcal{F}(\mathcal{D},S)$ and fix an element $(\Omega_0,\Gamma_0,\rho_0)\in \Omega^1(\mathcal{D},\mathfrak{s})\oplus\mathcal{F}(\mathcal{D},E)$. The choice $(\Omega_0,\Gamma_0,\rho_0)=(0,0,0)$ is allowed and important. Define the curve
\[
(\omega(t),\gamma(t)):=(\Lambda(t)^{-1}\dot\Lambda(t),\Lambda(t)^{-1}\dot r(t))\in\mathcal{F}(\mathcal{D},\mathfrak{s}).
\]
Let $(\Omega_0(t),\Gamma_0(t),\rho_0(t),u_0(t),w_0(t),f_0(t))$ and $(\Omega(t),\Gamma(t),\rho(t),u(t),w(t),f(t))$ be two curves related by the conditions
\[
(\Omega,\Gamma,\rho)=(\Lambda,r)^{-1}(\Omega_0,\Gamma_0,\rho_0)\quad\text{and}\quad (u,w,f)=(\Lambda,r)^{-1}(u_0,w_0,f_0)
\]
and such that
\[
\Omega_0(0)=\Omega_0, \quad \Gamma_0(0)=\Gamma_0,\quad \rho_0(0)=\rho_0.
\]
Suppose, for simplicity, that $(\Lambda(0),r(0))=(e,0)$. Then, the following are equivalent.

\begin{itemize}
\item[\bf{i}] Hamilton's variational principle
\[
\delta \int_{t_0}^{t_1} L_{(\Omega_0,\Gamma_0,\rho_0)}(\Lambda,\dot\Lambda,r,\dot r)dt=0,
\]
holds, for variations $\delta\Lambda(t),\delta r(t)$ vanishing at the endpoints.
\item[\bf{ii}] Hamilton's variational principle
\[
\delta \int_{t_0}^{t_1} \left(L(\Lambda,\dot\Lambda,r,\dot r,\Omega_0,\Gamma_0,\rho_0)+\int_\mathcal{D}\dot\Omega_0\cdot u_0+\dot\Gamma_0\!\cdot\! w_0+\dot\rho_0\!\cdot\!f_0\right)dt=0,
\]
holds, for variations $\delta\Lambda(t), \delta r_0(t), \delta \Omega_0(t), \delta\Gamma_0(t), \delta \rho_0(t), \delta u_0(t), \delta w_0(t)$, and $\delta f_0(t)$ vanishing at the endpoints.
\item[\bf{iii}] Hamilton's variational principle
\begin{align*}
\delta \int_{t_0}^{t_1}&\left(L\left(\Lambda,\dot\Lambda,r,\dot r,\operatorname{Ad}_\Lambda\Omega+\Lambda\mathbf{d}\Lambda^{-1},\Lambda\Gamma-(\operatorname{Ad}_\Lambda\Omega)r-\mathbf{d}^{\Lambda\mathbf{d}\Lambda^{-1}}r,\Lambda\rho-r\right)\phantom{\int_\mathcal{D}}\right.\\
&\quad+\left.\int_\mathcal{D} \left[ u\!\cdot\!\left(\dot\Omega-\mathbf{d}^\Omega\omega\right)+w\!\cdot\!\left(\dot\Gamma+\omega\Gamma-\mathbf{d}^\Omega\gamma\right)+f\!\cdot\!\left(\dot\rho+\omega\rho-\gamma\right)\right]\right),
\end{align*}
where $\omega=\Lambda^{-1}\dot\Lambda$ and $\gamma=\Lambda^{-1}\dot r$, holds, for variations $\delta\Lambda(t)$, $\delta r(t)$, $\delta \Omega(t)$, $\delta\Gamma(t)$, $\delta \rho(t)$, $\delta u(t)$, $\delta w(t)$, and $\delta f(t)$ vanishing at the endpoints.
\item[\bf{iv}] The reduced variational principle
\begin{align*}
&\delta \int_{t_0}^{t_1} \left(l(\omega,\gamma,\Omega,\Gamma,\rho) 
\phantom{\int_{\mathcal{D}}} \right.\\
&\qquad \left. +\int_\mathcal{D} \left[ u\!\cdot\!\left(\dot\Omega-\mathbf{d}^\Omega\omega\right)+w\!\cdot\!\left(\dot\Gamma+\omega\Gamma-\mathbf{d}^\Omega\gamma\right)+f\!\cdot\!\left(\dot\rho+\omega\rho-\gamma\right)\right]\right)dt=0,
\end{align*}
holds, for variations $\delta \Omega(t), \delta\Gamma(t), \delta \rho(t), \delta u(t), \delta w(t)$, and $\delta f(t)$ vanishing at the endpoints, and variations $\delta\omega(t)$ and $\delta\gamma(t)$ of the form
\[
\delta\omega=\frac{\partial\eta}{\partial t}+[\omega,\eta]\quad\text{and}\quad \delta\gamma=\frac{\partial v}{\partial t}+\omega v-\eta\gamma
\]
where $\eta(t), v(t)$ vanish at the endpoints.
\item[\bf{v}] The reduced variational principle
\[
\delta \int_{t_0}^{t_1}l(\omega,\gamma,\Omega,\Gamma,\rho)dt=0,
\]
holds, for variations $\delta\omega(t), \delta\gamma(t), \delta \Omega(t), \delta\Gamma(t), \delta \rho(t)$ of the form
\[
\delta\omega=\frac{\partial\eta}{\partial t}+[\omega,\eta]
,\quad 
\delta\gamma=\frac{\partial v}{\partial t}+\omega v-\eta\gamma
,
\]
\[
\delta\Omega=\mathbf{d}^\Omega\eta
,\quad
\delta\Gamma=\mathbf{d}^\Omega v-\eta\Gamma
\quad  \text{and}\quad
\delta\rho=v-\eta\rho,
\]
where $\eta(t), v(t)$ vanish at the endpoints.
\end{itemize}

Note that the curves $\Omega(t),\Gamma(t),\rho(t)$ are not present in part \textbf{i}, but they can be recovered from the curves $\Lambda(t),r(t)$, through the relation
\[
(\Omega(t),\Gamma(t),\rho(t))=\theta_{(\Lambda(t),r(t))^{-1}}(\Omega_0,\Gamma_0,\rho_0).
\]
The curves $\Omega_0(t)=\Omega_0,\Gamma_0(t)=\Gamma_0,\rho_0(t)=\rho_0$ are constant. The curves $u(t), w(t), f(t)$ are not present in parts $\textbf{i}$ and $\textbf{v}$, but they can be recovered by solving the equations
\[
\left\lbrace\begin{array}{l}
\displaystyle\vspace{0.2cm}
\dot u-\operatorname{ad}^*_\omega u+\gamma\diamond w-\frac{\delta l}{\delta \Omega}=0
,\\
\displaystyle\vspace{0.2cm}\dot w+\omega w-\frac{\delta l}{\delta \Gamma}=0
,\\
\displaystyle\dot f+\omega f-\frac{\delta l}{\delta \rho}=0
,
\end{array}\right.
\]
or, equivalently, the equations
\[
\left\lbrace\begin{array}{l}
\displaystyle\vspace{0.2cm}
\dot u-\operatorname{ad}^*_{\Lambda^{-1}\dot\Lambda}u+(\Lambda^{-1}\dot r)\diamond w-\operatorname{Ad}^*_\Lambda\left(\frac{\partial L}{\partial\Omega_0}-r\diamond \frac{\partial L}{\partial\Gamma_0}\right)=0
,\\
\displaystyle\vspace{0.2cm}\dot w+(\Lambda^{-1}\dot\Lambda) w-\Lambda^{-1}\frac{\partial L}{\partial\Gamma_0}=0
,\\
\displaystyle\dot f+(\Lambda^{-1}\dot\Lambda)f -\Lambda^{-1}\frac{\partial L}{\partial\rho_0}=0.
\end{array}\right.
\]
The curve $(u_0(t),w_0(t),f_0(t))$ is related to $(u(t),w(t),f(t))$ through the relation
\[
(u_0,w_0,f_0)=(\Lambda,r)(u,w,f)
\]
or it can be obtained directly by solving the equations
\[
\dot u_0-\frac{\partial L}{\partial\Omega_0}=0
,\quad \dot w_0-\frac{\partial L}{\partial\Gamma_0}=0
\quad\text{and}\quad \dot f_0-\frac{\partial L}{\partial\rho_0}=0.
\]
All of these variational principles are equivalent to the affine Euler-Poincar\'e equation \eqref{AEP_Kirchhoff_generalized} together with the equations for $\Omega, \Gamma$, and $\rho$.

\paragraph{The case of Kirchhoff's rod.} We now rewrite the points $\bf{iv}$ and $\bf{v}$ in the particular case $\mathcal{D}=[0,L]$ and $S=SE(3)$, that is, the case of Kirchhoff theory. We thus find that Kirchhoff's equations for a Lagrangian $l$ are equivalent to the following.
\begin{itemize}
\item[\bf{iv}] The reduced variational principle
\begin{align*}
&\delta \int_{t_0}^{t_1} \left(l(\boldsymbol{\omega},\boldsymbol{\gamma},\boldsymbol{\Omega},\boldsymbol{\Gamma},\boldsymbol{\rho})
+\int_\mathcal{D} \left[ \boldsymbol{u}\!\cdot\!\left(\dot{\boldsymbol{\Omega}}-\partial_s\boldsymbol{\omega}-\boldsymbol{\Omega}\times\boldsymbol{\omega}\right) \right.\phantom{\int_{\mathcal{D}}}\right.\\
&\qquad\qquad\left.\phantom{\int_{\mathcal{D}}}\left.+\boldsymbol{w}\!\cdot\!\left(\dot{\boldsymbol{\Gamma}}+\boldsymbol{\omega}\times\boldsymbol{\Gamma}-\partial_s\boldsymbol{\gamma}-\boldsymbol{\Omega}\times\boldsymbol{\gamma}\right)+\boldsymbol{f}\!\cdot\!\left(\dot{\boldsymbol{\rho}}+\boldsymbol{\omega}\times\boldsymbol{\rho}-\boldsymbol{\gamma}\right)\right]\right)dt=0,
\end{align*}
holds, for variations $\delta \boldsymbol{\Omega}$, $\delta\boldsymbol{\Gamma}$, $\delta \boldsymbol{\rho}$, $\delta \boldsymbol{u}$, $\delta \boldsymbol{w}$, and $\delta \boldsymbol{f}$ vanishing at the endpoints, and variations $\delta\boldsymbol{\omega}$ and $\delta\boldsymbol{\gamma}$ of the form
\[
\delta\boldsymbol{\omega}=\frac{\partial\boldsymbol{\eta}}{\partial t}+\boldsymbol{\omega}\times\boldsymbol{\eta}\quad\text{and}\quad \delta\boldsymbol{\gamma}=\frac{\partial \boldsymbol{v}}{\partial t}+\boldsymbol{\omega}\times \boldsymbol{v}-\boldsymbol{\eta}\times\boldsymbol{\gamma}
\]
where $\boldsymbol{\eta}(t), \boldsymbol{v}(t)$ vanish at the endpoints.
\item[\bf{v}] The reduced variational principle
\[
\delta \int_{t_0}^{t_1}l(\boldsymbol{\omega},\boldsymbol{\gamma},\boldsymbol{\Omega},\boldsymbol{\Gamma,\rho})dt=0,
\]
holds, for variations $\delta\boldsymbol{\omega}, \delta\boldsymbol{\gamma}, \delta \boldsymbol{\Omega}, \delta\boldsymbol{\Gamma}, \delta \boldsymbol{\rho}$ of the form
\[
\delta\boldsymbol{\omega}=\frac{\partial\boldsymbol{\eta}}{\partial t}+\boldsymbol{\omega}\times\boldsymbol{\eta}
,\quad 
\delta\boldsymbol{\gamma}=\frac{\partial \boldsymbol{v}}{\partial t}+\boldsymbol{\omega}\times \boldsymbol{v}-\boldsymbol{\eta}\times\boldsymbol{\gamma}
,
\]
\[
\delta\boldsymbol{\Omega}=\partial_s\boldsymbol{\eta}+\boldsymbol{\Omega}\times\boldsymbol{\eta}
,
\quad\delta\boldsymbol{\Gamma}=\partial_s\boldsymbol{v}+\boldsymbol{\Omega}\times\boldsymbol{v}
+ \boldsymbol{\Gamma}\times \boldsymbol{\eta}
\quad  \text{and}
\quad\delta\boldsymbol{\rho}=\boldsymbol{v}-\boldsymbol{\eta}\times\boldsymbol{\rho}
,
\]
where $\boldsymbol{\eta}(t), \boldsymbol{v}(t)$ vanish at the endpoints.
\end{itemize}

\noindent
The variables $(\boldsymbol{u},\boldsymbol{w},\boldsymbol{f})$ are conjugate momenta to the constraint variables $(\boldsymbol{\Omega},\boldsymbol{\Gamma},\boldsymbol{\rho})$, respectively, in the reduced variational principle in point {\bf iv}. 

Recall from the general theory developed in Section \ref{sec_AEP}, that the integrand appearing in the  action principle in \textbf{iv} above is the reduced Lagrangian $l^V$ associated to the cotangent bundle Lagrangian $L^V$, by standard Lagrangian reduction $TQ\rightarrow TQ/G$, see \eqref{def_L_V} and \eqref{def_l_V}, for the definition of these Lagrangians. In particular, \eqref{def_L_V} shows how one can construct $L^V$ from the Lagrangian $L=L(\Lambda,\dot\Lambda,\boldsymbol{r},\dot{\boldsymbol{r}})$ of the Kirchhoff rod.

{\footnotesize

\bibliographystyle{new}

}

\end{document}